\documentclass[aps,nofootinbib,prd]{revtex4}

\usepackage{lscape}
\usepackage{epsfig,epsf}
\usepackage{amsmath}
\usepackage{amsthm}
\usepackage{amsfonts}
\usepackage{amssymb}
\usepackage{dsfont}
\usepackage{multirow}
\usepackage{appendix}
\usepackage{slashed}
\usepackage[active]{srcltx}
\usepackage{psfrag}
\usepackage{subfigure}
\usepackage[colorlinks,citecolor=blue,urlcolor=red,linkcolor=red]{hyperref}
\usepackage{multirow}

\begin{document}

\title{{\Large{\bf  Semileptonic $B_{s}\to K_0^*(1430)$ transitions with the light-cone sum rules}}}

\author{\small
R. Khosravi\footnote {e-mail: rezakhosravi @iut.ac.ir }}

\affiliation{\emph{Department of Physics, Isfahan University of
Technology, Isfahan 84156-83111, Iran \\ Department of Physics,
Shiraz University, Shiraz 71454, Iran}}

\begin{abstract}
In the $\rm SU(3)_F$ symmetry limit, using two- and three-particle
distribution amplitudes of $B^{\pm}$-meson for $B_s$,  the
transition form factors of semileptonic $B_{s}\to K_{0}^*(1430)$
decays are calculated in the framework of the light-cone sum rules.
The two-particle distribution amplitudes, $\varphi_{_{+}}(\omega)$
and $\varphi_{_{-}}(\omega)$ have the most important contribution in
estimation of the form factors $f_{+}(q^2)$, $f_{-}(q^2)$ and
$f_{T}(q^2)$. The knowledge of the behavior of
$\varphi_{_+}(\omega)$ is still rather limited. Therefore, we
consider three different parametrizations for the shapes of
$\varphi_{_+}(\omega)$ that are derived from the phenomenological
models. Using the form factors $f_{+}$, $f_{-}$ and $f_{T}$, the
semileptonic $B_s \to K_0^*(1430) l \bar{\nu_{l}}$ and $B_s \to
K_0^*(1430) l \bar{l}/\nu \bar{\nu}$, $l=e, \mu, \tau$ decays are
analyzed. The branching fractions for the aforementioned decays, in
addition the longitudinal lepton polarization asymmetries are
calculated. A comparison between our results with predictions of
other approaches is provided.
\end{abstract}


\maketitle

\section{Introduction}
The scalar meson is a meson with total spin $0$ and even parity.
They are often produced in proton-antiproton annihilation, decays of
heavy flavor mesons, meson-meson scattering, and  radiative decays
of vector mesons. Among the scalar mesons, study of the light scalar
mesons up to $1.5\, \mbox{GeV}$ is important because their quark
content is still a common problem for high energy physics and may be
explained in a number of different ways, for example, considering as
a meson-meson molecules state \cite{molecules} or as a tetraquark
multiplet \cite{tetraquark}.

According to the quark model, the scalar mesons about $1\,
\mbox{GeV}$ are arranged into two $SU(3)$ nonets, in two scenarios:

Scenario 1 (S1): the light scalar mesons are assumed to compose from
two quarks. The nonet mesons below $1\, \mbox{GeV}$ are treated as
the lowest lying states, and the nonet mesons near $1.5\,
\mbox{GeV}$ are the excited states corresponding to the lowest lying
states.

Scenario 2 (S2): the nonet mesons below $1\, \mbox{GeV}$ may be
considered as four-quark bound states, and the other nonet mesons
are composed from two quarks and viewed as the lowest lying states.

Both scenarios in quark model agree that $K^*_0(1430)$ with the mass
of greater than $1\, \mbox{GeV}$ is a scalar meson with two quarks
dominated by the $s\bar u$ or $s \bar d$ state. However in S1, it is
regarded as an excited state, and in S2, it is seen as a ground
state. In the framework of the light-cone sum rules (LCSR),
differences between $K^*_0(1430)$ states in the two scenarios are
applied through different distribution amplitudes (DA's) and decay
constants \cite{CheChuYan}.

In this paper, our aim is to consider the semileptonic transitions
of $B_{s}$ to $K_{0}^*(1430)$ in the LCSR using the $B_s$-meson
DA's. In usual, the LCSR method is applied to calculate the form
factors of the heavy-to-light decays by utilizing the light meson
DA's. For this purpose, two-point correlation function is written
based on the light meson. Therefore, light-cone distribution
amplitudes (LCDA's) of the light meson appear in theoretical
calculations of the correlation function
\cite{Ruckl,Simma,Belyaev,Weinzierl,BBraun,Bagan,Ball,Zwicky,BZwicky}.
The LCDA's of the light mesons are related to the dynamics of
partons in long distance. Still, there is very limited knowledge of
the nonperturbative parameters determining these LCDA's. In the case
of the light scalar mesons, including $K_0^*(1430)$, this problem is
twofold because their internal structures are basically unknown. For
this reason, it is necessary to use a method of calculation that is
independent of the DA's of the scalar mesons.

In a new approach to the LCSR method related to the semileptonic $B$
decays, it was proposed to insert the correlation function between
vacuum and $B$-meson \cite{Offen}. In this technique the so-called
soft or endpoint, the correlation function is expanded in terms of
the DA's of $B$-meson, near the light-cone region \cite{KMO,BrKhod}.
Therefore, the transition form factors for exclusive decays of $B$
to light mesons are connected to the DA's that depend on the
dynamical information of $B$-meson. Two-particle DA of $B$-meson,
$\varphi_{_+}(\omega)$ plays a particularly prominent role in this
new approach to exclusive semileptonic decays. The knowledge of the
behavior of $\varphi_{_+}(\omega)$ is still rather limited due to
the poor understanding of nonperturbative QCD dynamics (for
instance, see Refs. \cite{BraManash,GalNeu,XuZhao,ZhaRad}). So far,
several models for the shape of $\varphi_{_+}(\omega)$ have been
proposed based on the QCD sum rules (QCDSR) \cite{Grozin,Braun}, the
LCSR
\cite{GenonSachrajda,Kou,LeeNeubert,BellFeldmann,FeldLange,WangShen,BeneBraun},
and the QCD factorization \cite{JiMan}. Also, the functional form of
the three-particle $B$-meson DA's have been estimated in several
models \cite{KMO,JiMan,ShenWangWei}.

In this work, the form factors of the semileptonic $B_s \to
K_0^*(1430)$ transitions are investigated in the new approach of the
LCSR with the two- and three-particle DA's of $B_s$-meson in the
$\rm SU(3)_F$ symmetry limit. Utilizing these form factors, the
semileptonic $B_s \to K_0^*(1430) l \bar{\nu_{l}}$ and $B_s \to
K_0^*(1430) l \bar{l}/\nu \bar{\nu}$, $l=e, \mu, \tau$ decays are
analyzed. In the standard model (SM), the rare semileptonic $B_s \to
K_0^*(1430) l \bar{l}$ decays occur at loop level instead of tree
level, by electroweak penguin and weak box diagrams via the flavor
changing neutral current (FCNC) transitions of $b\to d l^+ l^-$ at
quark level. In particle physics, reliable considering of the FCNC
decays of $B$-meson is very important since they are sensitive to
new physics (NP) contributions to penguin operators. So, to test the
SM and look for NP, we need to determine the SM predictions for FCNC
decays and compare these results to the corresponding experimental
values.

This work is organized as follows: In Sec II, according to the
effective weak Hamiltonian of the FCNC transition $b \to d~ l^+
l^-$, the form factors of the semileptonic $B_{s}\to K_{0}^*$ decays
are calculated with the LCSR model using the $B_s$-meson DA's. These
form factors are basic parameters in studying the forward-backward
asymmetry, longitudinal lepton polarization asymmetry and branching
fraction of semileptonic decays. Our numerical and analytical
results and their comparison with the predictions of other
approaches are presented in Sec III. The last section is dedicated
to conclusion. Future experimental measurement can give valuable
information about these aforesaid decays and the nature of the
scalar meson $K_0^*(1430)$.

\section{$B_{s}\to K_{0}^*\, l^{+} l^{-}$  Form Factors with the LCSR}

According to the effective weak Hamiltonian of the $b \to d~ l^+
l^-$ transition presented in Appendix, the matrix element for the
FCNC decay $b \to d$ can be written as:
\begin{eqnarray}\label{eq1}
{\cal M}= \frac{G_{F}\alpha}{2\sqrt{2}\pi} V_{tb}V_{td}^{*}\Bigg[
C_9^{\rm eff} \,\, \bar {d} \gamma_\mu (1-\gamma_5) b~ \overline {l
}\gamma_\mu  l + C_{10}\,\, \bar {d} \gamma_\mu (1-\gamma_5) b~
\overline {l } \gamma_\mu \gamma_5 \l - 2\, C_7^{\rm
eff}\,\,\frac{m_b}{q^2}\, \bar {d} ~i\sigma_{\mu\nu} q^\nu
(1+\gamma_5) b~  \overline {l}  \gamma_\mu l \Bigg],
\end{eqnarray}
where $G_F$ is the Fermi constant, $\alpha$  is the fine structure
constant at $Z$ mass scale, and $V_{ij}$ are elements of the
Cabbibo- Kobayashi-Maskawa (CKM) matrix. $\bar d \gamma_\mu
(1-\gamma_5) b$ and $\bar d \sigma_{\mu\nu}q^\nu (1+\gamma_5) b$ are
the transition currents denoted with $J^{V-A}_{\mu}$ and
$J^{T}_{\mu}$, respectively. This decay amplitude also contains two
effective Wilson coefficients $C_7^{\rm eff}$ and $C_9^{\rm eff}$,
where $C_7^{\rm eff}= C_7-C_5/3-C_6$ and $C_9^{\rm eff}$ is
explained in Appendix.

\begin{figure}[th]
\begin{center}
\includegraphics[width=4cm,height=2cm]{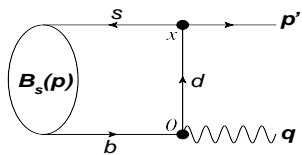}
\caption{leading-order diagram for $B_s \rightarrow K_0^* l^+ l^-$
decays.} \label{F1}
\end{center}
\end{figure}

To investigate the form factors of $B_s \to K_{0}^* l^{+} l^{-}$
decays via the LCSR, the two-point correlation functions are
constructed from the transition currents $J_{\mu}^{V-A(T)}$, and
interpolating current $J^{K_0^*}$ of the scalar meson $K_0^*$,
inserted between vacuum and $B_s$-meson as follows:
\begin{eqnarray} \label{eq2}
\Pi_{\mu}^{V-A(T)}(p^\prime,q)=i\int {d^{4}x}~ e^{ip^\prime.x}
\langle0| T\left\{
J^{K_0^*}(x)J_{\mu}^{V-A(T)}(0)\right\}|B_s(p)\rangle,
\end{eqnarray}
where $T$ is the time ordering operator, $J^{K_0^*}=\bar{s}(x) d(x)
$ and $q=p-p^\prime$. The external momenta $p^\prime$ and $q$ are
related to the interpolating and transition currents, $J^{K_0^*}$
and $J_{\mu}^{V-A(T)}$ respectively, so that $p^2= (p^\prime + q)^2
= m^2_B$. The leading-order diagram for $B_s \rightarrow K_0^* l^+
l^-$  decays is depicted in Fig. \ref{F1}.

The correlation functions in Eq. (\ref{eq2}) are  complex quantities
and have two aspects: phenomenological and theoretical.  Hadronic
parameters like form factors appear in the phenomenological or
physical representation of the correlation functions. The
theoretical or QCD side of the correlation functions is obtained in
terms of the DA's of $B_s$-meson. Equating coefficients of the
corresponding lorentz structures from both representations through
the dispersion relation,
\begin{eqnarray}\label{eq3}
\Pi_{\mu}(p^\prime,q)=\frac{1}{\pi}\int_{0}^{\infty} ds
\frac{\mbox{Im}\,\Pi_{\mu}(s)}{s-p'^2}\,,
\end{eqnarray}
and applying Borel transformation to suppress the contributions of
the higher states and continuum, the form factors are calculated
from the LCSR.

Inserting a complete set of intermediate states with the same
quantum number as the interpolating current $J^{K_0^*}$, in Eq.
(\ref{eq2}), and isolating the pole term of the lowest scalar meson
$K_0^*$, and then applying Fourier transformation, the
phenomenological representations of the correlation functions are
obtained, as follows:
\begin{eqnarray} \label{eq4}
\Pi_{\mu}^{V-A(T)}(p^\prime,q)&=&\frac{1}{m_{K_0^*}^2-p^{\prime2}}
\langle 0 | J^{K_0^*}(p^\prime) |K_0^*(p^\prime) \rangle \langle
K_0^*(p^\prime) | J^{V-A(T)}_{\mu}| B_s(p) \rangle \nonumber\\
&+&\sum_{h}\,\frac{1}{m_{h}^2-p^{\prime2}}\langle 0 |
J^{K_0^*}(p^\prime) |h(p^\prime) \rangle \langle h(p^\prime) |
J^{V-A(T)}_{\mu}| B_s(p) \rangle\,.
\end{eqnarray}
To continue, we define the spectral density functions of higher
resonances and the continuum of states as
\begin{eqnarray} \label{eq5}
\rho^{h, V-A(T)}_{\mu}(s)\equiv\pi\,\sum_{h}\,\langle 0 |
J^{K_0^*}(p^\prime) |h(p^\prime) \rangle \langle h(p^\prime) |
J^{V-A(T)}_{\mu}| B_s(p) \rangle \, \delta(s-m^{2}_{h}).
\end{eqnarray}
Inserting the spectral density functions in Eq. (\ref{eq4}), the
correlation functions are obtained as
\begin{eqnarray}\label{eq6}
\Pi^{V-A(T)}_{\mu} (p^\prime,q) &=&
\frac{1}{m_{K_0^*}^2-p^{\prime2}} \langle 0 | J^{K_0^*}(p^\prime)
|K_0^*(p^\prime) \rangle \langle K_0^*(p^\prime) | J^{V-A(T)}_{\mu}|
B_s(p) \rangle + \frac{1}{\pi}\int_{s_0}^{\infty} ds
\frac{\rho_{\mu}^{h,{V-A(T)}}(s)}{s-p'^2},
\end{eqnarray}
where $s_0$ is the continuum threshold of $K_0^*$ meson. The matrix
element, $\langle 0|J^{K_0^*}|K_0^*\rangle =f_{K_0^*} m_{K_0^*}$,
where $f_{K_0^*}$ is the leptonic decay constant of the scalar meson
$K_0^*$. Considering parity and using Lorentz invariance, the
transition matrix elements, $\langle K_0^*(p^\prime) |
J^{V-A(T)}_{\mu}| B_s(p) \rangle$, can be parametrized as:
\begin{eqnarray}\label{eq7}
\langle K_0^*(p^\prime) |
J^{V-A}_{\mu}| B_s(p) \rangle&=& i\left[P_{\mu}f_+(q^2)\, +\,q_\mu  f_-(q^2) \right],\nonumber\\
\langle K_0^*(p^\prime) | J^{T}_{\mu}| B_s(p) \rangle&=&-
\frac{1}{m_{B_s} + m_{K_0^*}} \left[ P_{\mu} q^2 -q_\mu(m_{B_s}^2 -
m_{K_0^*}^2) \right]f_T(q^2),
\end{eqnarray}
where $f_+(q^2)$, $f_-(q^2)$ and $f_T(q^2)$\, are the transition
form factors, which only depend on  the momentum transfer squared
$q^2$, $P_{\mu}=(p^\prime +p)_\mu$, and $q_\mu = (p- p^\prime)_\mu$.
Substituting Eq. (\ref{eq7}) in Eq. (\ref{eq6}),  we obtain
\begin{eqnarray}\label{eq8}
\Pi^{V-A}_{\mu} (p^\prime,q)&=& if_{K_0^*}
m_{K_0^*}\left[\frac{P_{\mu}f_+(q^2)\, +\,q_\mu
f_-(q^2)}{m_{K_0^*}^2-p^{\prime2}}\right]+
\frac{1}{\pi}\int_{s_0}^{\infty} ds
\frac{\rho_{\mu}^{h,{V-A}}(s)}{s-p'^2}\,,\nonumber\\
\Pi^{T}_{\mu} (p^\prime,q)&=& -\frac{f_{K_0^*} m_{K_0^*} }{m_{B_s} +
m_{K_0^*}} \left[\frac{P_{\mu} q^2 -q_\mu(m_{B_s}^2 -
m_{K_0^*}^2)}{m_{K_0^*}^2-p^{\prime2}}\right]f_T(q^2)+
\frac{1}{\pi}\int_{s_0}^{\infty} ds
\frac{\rho_{\mu}^{h,{T}}(s)}{s-p'^2}\,.
\end{eqnarray}

To extract the theoretical or QCD side, the correlation functions in
Eq. (\ref{eq2}) are expanded in the limit of large $m_b$ in heavy
quark effective theory (HQET). In the HQET, the relation between the
momentum and four-velocity of $B_s$-meson is as: $p = m_b v+k$,
where  $k$ is the residual momentum. Using the relation $p=q+p'$ and
$p = m_b v+k$, the four-momentum transfer $\widetilde{q}$ is defined
as: $k-p'=q - m_b v\equiv\widetilde{q}$, where $\widetilde{q}$ is
called static part of $q$. Up to $1/m_b$ corrections, the
$B_s$-meson state can be estimated by the relativistic normalization
of it $|B_s(p)\rangle = |B_s (v)\rangle$, and the correlation
functions $\Pi_{\mu}^{V-A(T)}(p ^\prime,q)$ can be approximated to
$\widetilde{\Pi}_{\mu}^{V-A (T)}(p^\prime,\widetilde{q})$,
\begin{eqnarray}\label{eq9}
\Pi_{\mu}^{V-A(T)}(p ^\prime,q)&=&\widetilde{\Pi}_{\mu}^{V-A
(T)}(p^\prime,\widetilde{q})+\mathcal{O}(1/m_b).
\end{eqnarray}
Also, the $b$-quark field is substituted by the effective field as
$b(x) = e^{-im_b v x}h_v(x)$. Therefore, the correlation functions
in the heavy quark limit, ($m_b \to \infty$), become  \cite{KMO}:
\begin{eqnarray}\label{eq10}
\widetilde{\Pi}_{\mu}^{V-A }(p^\prime,\widetilde{q})&=& i\int d^4x~
e^{ip^\prime.x}\langle 0| T \{\bar{s}(x) S_d(x) \gamma_{\mu} (1-
\gamma_{5}) h_v(0)\}
|B_s(v)\rangle,\nonumber\\
\widetilde{\Pi}_{\mu}^{T}(p^\prime,\widetilde{q})&=& i\int d^4x~
e^{ip^\prime.x}\langle 0| T \{\bar{s}(x) S_d(x) \sigma_{\mu\nu}q^\nu
(1+\gamma_5)  h_v(0)\} |B_s(v)\rangle.
\end{eqnarray}
The full-quark propagator, $S_d(x)$ of a massless quark in the
external gluon field in the Fock-Schwinger gauge is as follows
\cite{BalBra}:
\begin{eqnarray}\label{eq11}
S_d(x) &=& i \int \frac{d^{4}k}{(2\pi)^{4}}
e^{-ik.x}\left\{\frac{\not\! k}{k^{2}}+\int^{1}_{0}du~
G_{\lambda\rho}(u x)\left[\frac{1}{k^2}
ux^{\lambda}\gamma^{\rho}-\frac{1}{2k^{4}}\not\! k
\sigma^{\lambda\rho}\right] \right\}.
\end{eqnarray}
When the full-quark propagator $S_d(x)$ in Eq. (\ref{eq11}) is
replaced in Eq. (\ref{eq10}), operators between vacuum mode and
$B_s(v)$-state create the nonzero matrix elements as $\langle 0|\bar
{s}{_{\alpha}(x){h_{v\beta}(0)}}|B_s(v)\rangle$ and $\langle
0|\bar{s}_\alpha(x) G_{\lambda\rho}(ux)
h_{v\beta}(0)|B_s(v)\rangle$. These matrix elements are obtained in
terms of two- and three-particle DA's of $B_s$-meson, as \cite{KMO}
\begin{eqnarray}\label{eq12}
\langle 0|\bar {s}{_{\alpha}(x){h_{v\beta}(0)}}|B_s(v)\rangle &=&
-\frac{i f_{B} m_{B}}{4} \int^{\infty}_{0} d\omega~ e^{-i\omega v.x}
\left\{(1+\not\! v) \left[\varphi_{_+}(\omega)-\frac{\not\! x}{2v.
x}(\varphi_{_+}(\omega)-\varphi_{_-}(\omega))\right]  \gamma_{5}
\right\}_{\beta\alpha},
\nonumber\\
\langle 0|\bar{s}_\alpha(x) G_{\lambda\rho}(ux)
h_{v\beta}(0)|B_s(v)\rangle &=& \frac{f_B m_B}{4}\int_0^\infty
d\omega \int_0^\infty d\xi  e^{-i(\omega+u\xi) v .x} \Big\{ (1 +
\not\!v)\Big[(v_\lambda\gamma_\rho-v_\rho\gamma_\lambda)
\left(\Psi_{_A}(\omega, \xi)-\Psi_{_V}(\omega, \xi)\right)\nonumber\\
&&-i\sigma_{\lambda\rho}\Psi_{_V}(\omega, \xi) -\frac{x_\lambda
v_\rho-x_\rho v_\lambda}{v. x}\,X_{_A}(\omega, \xi)+\frac{x_\lambda
\gamma_\rho-x_\rho \gamma_\lambda}{v. x}\,Y_{_A}(\omega,
\xi)\Big]\gamma_5\Big\}_{\beta\alpha},
\end{eqnarray}
where $\varphi_{_+}$ and $\varphi_{_-}$ are the two-particle DA's
and $\Psi_{_A}$, $\Psi_{_V}$, $X_{_A}$ and $Y_{_A}$ are four
independent three-particle DA's of $B_s$-meson.

To calculate the correlation functions in terms of the two- and
three-particle DA's,  we substitute Eq. (\ref{eq12}) in the matrix
elements $\langle 0|\bar
{s}{_{\alpha}(x){h_{v\beta}(0)}}|B_s(v)\rangle$ and $\langle
0|\bar{s}_\alpha(x) G_{\lambda\rho}(ux) h_{v\beta}(0)|B_s(v)\rangle$
that appear in the correlation functions and then the integrals are
investigated. Generally, the results of the calculations can be
arranged in the following form:
\begin{eqnarray}\label{eq013}
\widetilde{\Pi}_{\mu}^{V-A} (p^\prime,\widetilde{q})&=&i\,\left[
\widetilde{\Pi}_{+}(p^\prime,\widetilde{q})\,P_{\mu}+\widetilde{\Pi}_{-}(p^\prime,\widetilde{q})\,q_{\mu}\right]\,,\nonumber\\
\widetilde{\Pi}_{\mu}^{T}
(p^\prime,\widetilde{q})&=&\widetilde{\Pi}_{T}(p^\prime,\widetilde{q})\,P_{\mu}
+...\,,
\end{eqnarray}
and $\widetilde{\Pi}_{+}$, $\widetilde{\Pi}_{-}$, and
$\widetilde{\Pi}_{T}$ are presented as follows:
\begin{eqnarray}\label{eq13}
\widetilde{\Pi}_{j}
(p^\prime,\widetilde{q})=\frac{1}{\pi}\int_{0}^{\infty} \frac{
d\sigma}{{\textbf{s}}(\sigma)-p'^{2}}\,{g_{j}(\sigma)}\,,
\end{eqnarray}
where $j=+, -, T$. In this representation of the theoretical part of
the correlation functions, $\sigma=\omega/m_{B_s}$ is the
integration variable, $g_{j}(\sigma)$ is a function of $\sigma$ in
terms of the $B_s$-meson DA's, and ${\textbf{s}}(\sigma)$ is defined
as
\begin{eqnarray}\label{eq14}
{\textbf{s}}(\sigma)=\sigma\, m^2_{B_s}
-\frac{\sigma}{\bar{\sigma}}\,q^2\,,
\end{eqnarray}
where $\bar{\sigma}=1-\sigma$.

On the other hand, using the dispersion relation, the theoretical
part of the correlation functions $\widetilde{\Pi}_{j}$ can be
related to its imaginary part as
\begin{eqnarray} \label{eq15}
\widetilde{\Pi}_{j}(p^\prime,\widetilde{q})&=&
\frac{1}{\pi}\int_{0}^{\infty} ds
\,\frac{{\rm{Im}}\widetilde{\Pi}_{j}(s)}{s-p'^2}.
\end{eqnarray}
At large spacelike $p'^2$, the quark-hadron duality approximation is
employed as:
\begin{eqnarray} \label{eq16}
\frac{1}{\pi}\int_{s_0}^{\infty} ds
\,\frac{{\rho}_{j}(s)}{s-p'^2}&\simeq&
\frac{1}{\pi}\int_{s_0}^{\infty} ds
\,\frac{{\rm{Im}}\widetilde{\Pi}_{j}(s)}{s-p'^2},
\end{eqnarray}
where ${\rho}_{j}(s)$ is the functional part of the tensor
${\rho}^{h}_{\mu}$ so that an expression similar to Eq.
(\ref{eq013}) can be written for it. Using Eqs. (\ref{eq15}) and
(\ref{eq16}) in Eq. (\ref{eq8}), and equating the coefficients of
the Lorentz structures $P_{\mu}$ and $q_{\mu}$, leads to the
following result.
\begin{eqnarray}\label{eq17}
\frac{1}{\pi}\int^{s_0}_{0} ds
\frac{{\rm{Im}}\widetilde{\Pi}_{\pm}(s)}{s-p'^2}&=&\frac{f_{K_0^*}
m_{K_0^*}}{m_{K_0^*}^2-p^{\prime2}}\,f_{\pm}(q^2)\,,\nonumber\\
\frac{1}{\pi}\int^{s_0}_{0} ds
\frac{{\rm{Im}}\widetilde{\Pi}_{T}(s)}{s-p'^2}&=&- \frac{f_{K_0^*}
m_{K_0^*} }{m_{B_s} + m_{K_0^*}}
\frac{q^2\,f_T(q^2)}{m_{K_0^*}^2-p^{\prime2}}\,.
\end{eqnarray}
Finally, according to  Eq. (\ref{eq13}) and Eq. (\ref{eq15}), it can
be concluded that
\begin{eqnarray}\label{eq18}
\frac{1}{\pi}\int_{0}^{\sigma_0} \frac{
d\sigma}{{\textbf{s}}(\sigma)-p'^{2}}\,{g_{\pm}(\sigma)}&=&
\frac{f_{K_0^*} m_{K_0^*}}{m_{K_0^*}^2-p^{\prime2}}\,f_{\pm}(q^2)\,
,\nonumber\\
\frac{1}{\pi}\int_{0}^{\sigma_0} \frac{
d\sigma}{{\textbf{s}}(\sigma)-p'^{2}}\,{g_{T}(\sigma)}&=&
-\frac{f_{K_0^*} m_{K_0^*} }{m_{K_0^*}^2-p^{\prime2}}
\,\frac{q^2\,f_T(q^2) }{m_{B_s} + m_{K_0^*}}\,.
\end{eqnarray}
To determine the effective threshold $\sigma_0$, the continuum
threshold of $K_0^*$ meson  $s_0$ is replaced in Eq. (\ref{eq14})
instead of ${\textbf{s}}$. A quadratic equation is created based on
the variable $\sigma$. By solving this equation, the value of
$\sigma_0$ is determined as follows:
\begin{eqnarray}\label{eq19}
\sigma_0=\frac{s_0+m_{B_s}^2-q^2-\sqrt{(s_0+m_{B_s}^2-q^2)^2-4s_0m_{B_s}^2}}{2m_{B_s}^2}.
\end{eqnarray}

Applying Borel transformation with respect to the variable $p'^{2}$
as:
\begin{eqnarray}\label{eq21}
B_{p^{\prime2}}(M^{2})(\frac{1}{p^{\prime2}-m^{2}})^{n}&=&\frac{(-1)^{n}}{\Gamma(n)}\frac{e^{-\frac{m^{2}}{M^{2}}}}{(M^{2})^{n}},
\end{eqnarray}
in Eq. (\ref{eq18}) in order to suppress the contributions of the
higher states, the form factors are obtained via the LCSR in terms
of the two- and three-particle DA's of $B_s$-meson. Our results for
$f_{+}(q^2)$, $f_{-}(q^2)$,  and $f_{T}(q^2)$ are presented as:
\begin{eqnarray}\label{eq22}
f_{+}(q^{2})&=&\frac{f_{B_s}m_{B_s}^2}{2
f_{K_0^*}m_{K_0^*}}e^{\frac{m_{K_0^*}^2}{M^2}}
\int_{0}^{\sigma_{0}}d\sigma\,e^{-\frac{{\textbf{s}}(\sigma)}{M^2}}
\left\{\varphi_{_+}(\sigma
m_{_{B_s}})-\frac{\widetilde{\varphi}_{_+}(\sigma
m_{_{B_s}})-\widetilde{\varphi}_{_-}(\sigma m_{_{B_s}})}{
\bar{\sigma} m_{_{B_s}} }\right.\nonumber\\
&+&\left. \int_{0}^{\sigma m_{B_s}}d\omega \int_{\sigma
m_{B_s}-\omega}^{\infty }\frac{d\xi}{\xi}
\left\{\left[(2u+2)\left(\frac{q^2-\bar{\sigma}^2
m_{B_s}^2}{\bar{\sigma}^3
M^2}+\frac{1}{\bar{\sigma}^2}\right)+\frac{(2u+1)m_{B_s}^2}{\bar{\sigma}
M^2}\right] \frac{\Psi_{_A}(\omega, \xi)-\Psi_{_V}(\omega,
\xi)}{m_{B_s}^2}\right.\right.\nonumber\\
&+&\left.\left.\frac{6u}{\bar{\sigma}M^2}\,\Psi_{_V}(\omega,
\xi)+\left[(2u-1)\left(\frac{q^2-\bar{\sigma}^2
m_{B_s}^2}{\bar{\sigma}^3 M^4}+\frac{1}{{\bar{\sigma}}^2
M^2}\right)+\frac{3}{{\bar{\sigma}^2}
M^2}\right]\frac{\widetilde{X}_{_A}(\omega,\xi)}{m_{B_s}}-\frac{4(u+3)}{\bar{\sigma}^2
M^2}\,
\frac{\widetilde{Y}_{_A}(\omega,\xi)}{m_{B_s}}\right\}\right\},
\nonumber\\ \nonumber\\
f_{-}(q^{2})&=&-\frac{f_{B_s}m_{B_s}^2}{2
f_{K_0^*}m_{K_0^*}}e^{\frac{m_{K_0^*}^2}{M^2}}
\int_{0}^{\sigma_{0}}d\sigma\,e^{-\frac{{\textbf{s}}(\sigma)}{M^2}}
\left\{\frac{(1+\sigma)}{\bar{\sigma}} \varphi_{_+}(\sigma
m_{_{B_s}})+\frac{\widetilde{\varphi}_{_+}(\sigma m_{_{B_s}})
-\widetilde{\varphi}_{_-}(\sigma m_{_{B_s}})}{
\bar{\sigma} m_{_{B_s}} }\right.\nonumber\\
&-&\left. \int_{0}^{\sigma m_{B_s}}d\omega \int_{\sigma
m_{B_s}-\omega}^{\infty }\frac{d\xi}{\xi}
\left\{\left[(2u+2)\left(\frac{q^2-\bar{\sigma}^2
m_{B_s}^2}{\bar{\sigma}^3
M^2}+\frac{1}{\bar{\sigma}^2}\right)-\frac{(2u+1) (1+\sigma)
m_{B_s}^2)}{\bar{\sigma}^2 M^2}\right] \frac{\Psi_{_A}(\omega,
\xi)-\Psi_{_V}(\omega,
\xi)}{m_{B_s}^2}\right.\right.\nonumber\\
&+&\left.\left.\frac{6u(1+\sigma)}{\bar{\sigma}^2
M^2}\,\Psi_{_V}(\omega,
\xi)+\left[\frac{(2u-1)(1+\sigma)}{\bar{\sigma}}\left(\frac{q^2-\bar{\sigma}^2
m_{B_s}^2}{\bar{\sigma}^3 M^4}+\frac{1}{{\bar{\sigma}}^2
M^2}\right)+\frac{4(u+\bar{\sigma})}{{\bar{\sigma}^3}
M^2}\right] \frac{\widetilde{X}_{_A}(\omega,\xi)}{m_{B_s}}\right.\right.\nonumber\\
&+&\left.\left.\frac{4(u+3)}{\bar{\sigma}^2 M^2}\,
\frac{\widetilde{Y}_{_A}(\omega,\xi)}{m_{B_s}}\right\}\right\},
\nonumber\nonumber\\
f_{T}(q^{2})&=&\frac{f_{B_s}m_{B_s}(m_{B_s}
+{m_{K_0^*}})}{2f_{K_0^*}m_{K_0^*}}e^\frac{m_{K_0^*}^2}{M^2}\int_{0}^{\sigma_{0}}d\sigma\,e^{-\frac{{\textbf{s}}(\sigma)}{M^2}}
\left\{\frac{\varphi_{_+}(\sigma m_{_{B_s}})}{\bar{\sigma}}\,
+\int_{0}^{\sigma m_{B_s}}d\omega \int_{\sigma
m_{B_s}-\omega}^{\infty }\frac{d\xi}{\xi}
\left\{\left[\frac{6u}{\bar{\sigma}^2 M^2}\right] \Psi_{_V}(\omega,
\xi)\right.\right.\nonumber\\
&+&\left.\left. \left[\frac{2u+1}{\bar{\sigma}^2 M^2}\right]
(\Psi_{_A}(\omega, \xi)-\Psi_{_V}(\omega,
\xi))-\left[\frac{q^2-\bar{\sigma}^2 m_{B_s}^2}{\bar{\sigma}^4
M^4}+2\right] \frac{\widetilde{X}_{_A} }{m_{B_s}}\right\}\right\},
\end{eqnarray}
where:
\begin{equation*}\label{eq23}
u = \frac{\sigma m_{B_s}-\omega}{\xi}, \qquad
\widetilde{\varphi}_{_{\pm}}(\sigma m_{B_s}) =  \int_0^{\sigma
m_{B_s}} d\tau {\varphi}_{_{\pm}}(\tau), \qquad
\widetilde{X}{_A}(\omega,\xi) =\int_0^\omega d\tau X_{_A}(\tau,\xi),
\qquad \widetilde{Y}{_A}(\omega,\xi) =\int_0^\omega d\tau
Y_{_A}(\tau,\xi).
\end{equation*}

\section{ Numerical Analysis }\label{sec3}
In this section, our numerical analysis of the form factors $f_{+}$,
$f_{-}$ and $f_{T}$ is presented for the semileptonic $B_{s}\to
K_0^*$ decays. The values are chosen for masses in $\mbox{GeV}$  as:
$m_{B_s}=5.37$, $m_{K_0^*}=(1.43\pm 0.05)$, $m_{\tau}=1.78$, and
$m_{\mu}=0.11$ \cite{PDG}. The leptonic decay constants are taken
as: $f_{K_0^*}=(427\pm 85)\,\mbox{MeV}$ \cite{Dong}, and
$f_{B_{s}}=(230.3 \pm 1.3)\, \mbox{MeV}$ \cite{Aoki}. Moreover, the
continuum threshold of $K_0^*$ meson, $s_0$ is equal to $(4.4 \pm
0.4) \mbox{GeV}^2$ \cite{Dong}. The values of the parameters
$\lambda _{E}^{2}$ and $\lambda _{H}^{2}$  of the $B_s$-meson DA's
are chosen as: $\lambda _{E}^{2}=(0.01 \pm 0.01) \,\mbox{GeV}^2$ and
$\lambda _{H}^{2}=(0.15 \pm 0.05) \,\mbox{GeV}^2$\cite{Rahimi}.

\begin{figure}[th]
\begin{center}
\includegraphics[width=5.5cm,height=5.5cm]{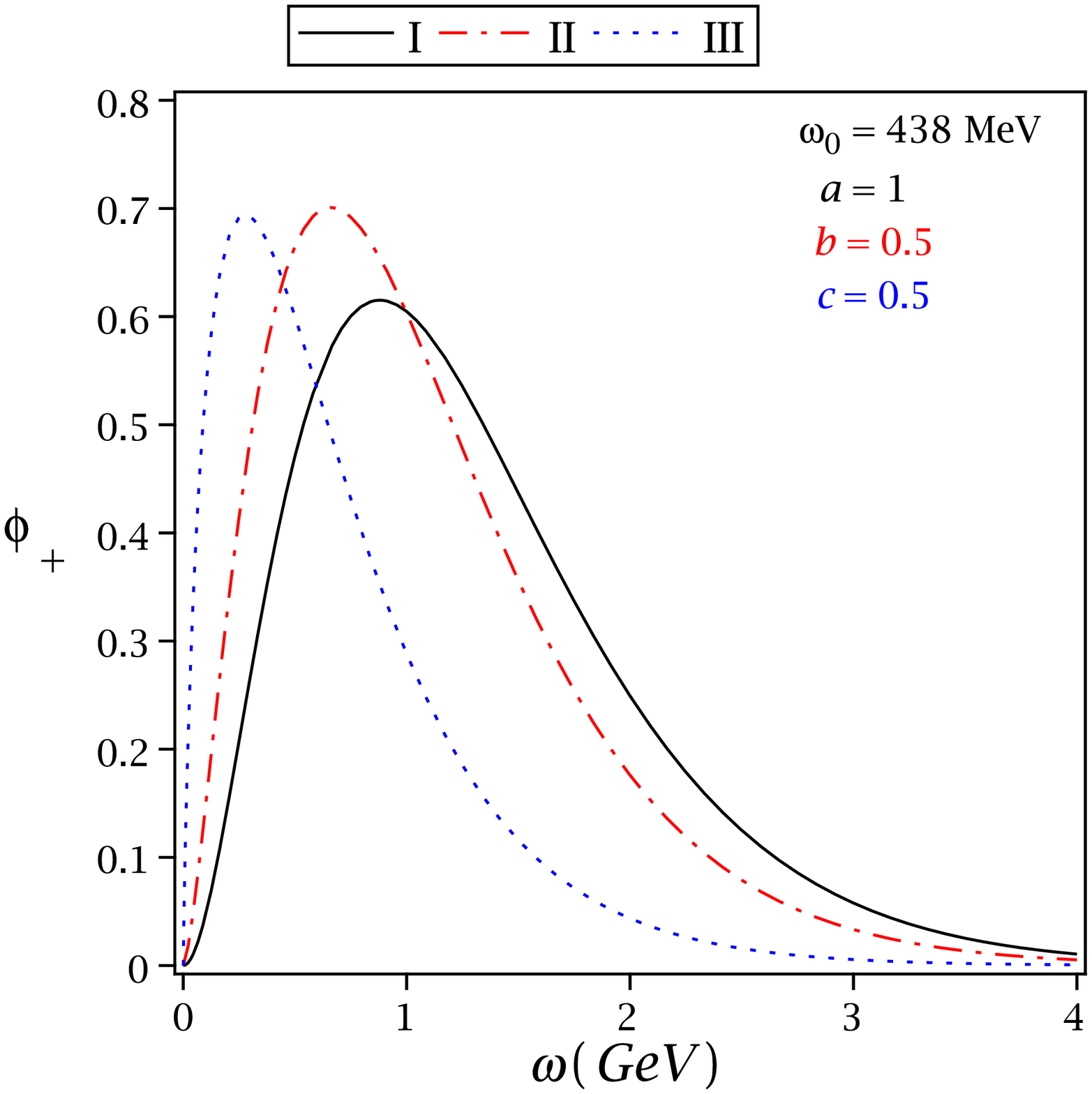}
\includegraphics[width=5.5cm,height=5.5cm]{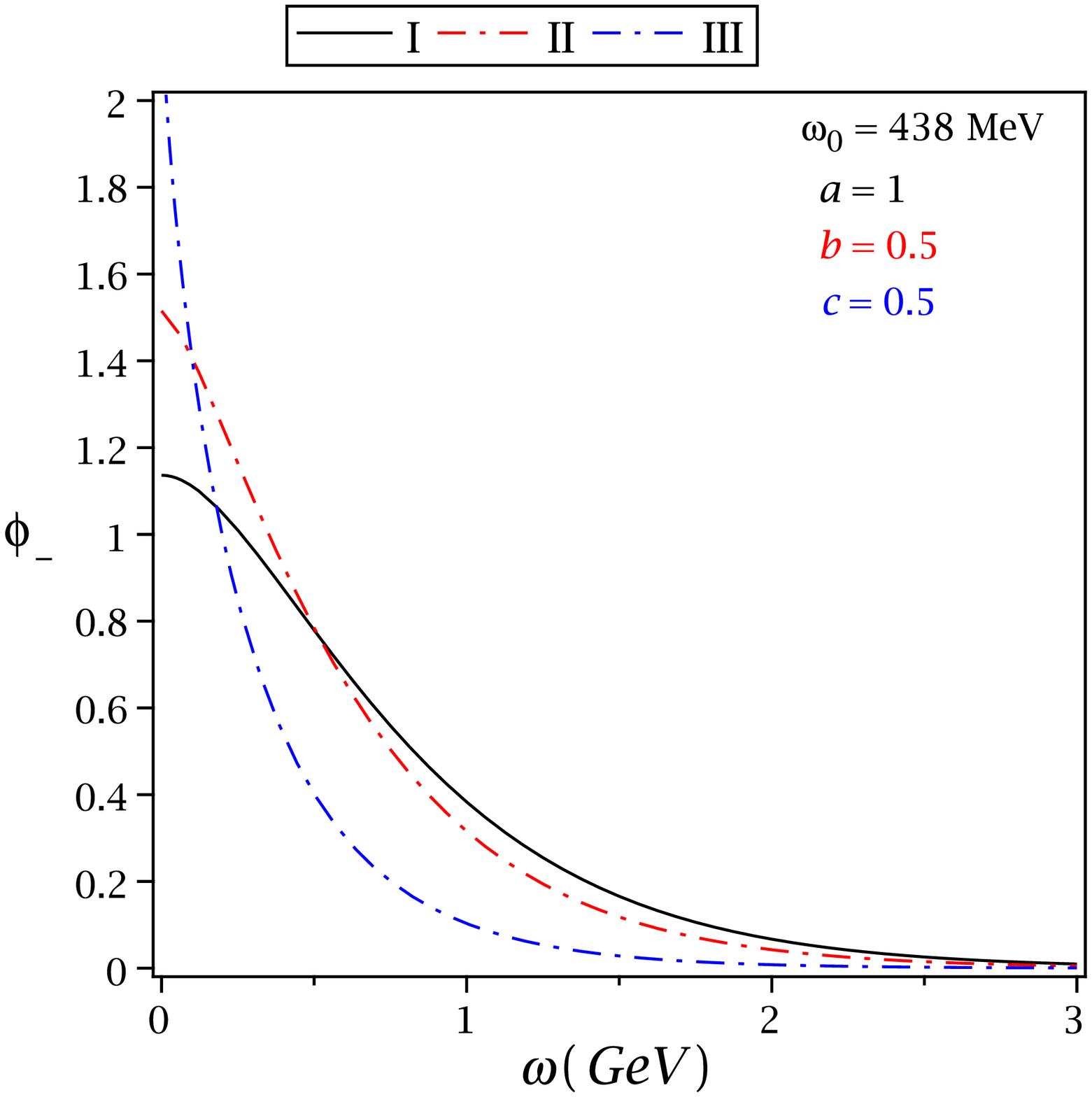}
\caption{The dependence of $\varphi_{_+}(\omega)$ and
$\varphi_{_-}(\omega)$ on $\omega$ for the three models.} \label{F2}
\end{center}
\end{figure}

The two-particle DA's of $B_s$-meson, $\varphi_{_{+}}(\omega)$ and
$\varphi_{_{-}}(\omega)$ have the most important contribution in
estimation of the form factors $f_{+}$, $f_{-}$, and $f_{T}$. The
knowledge of the behavior of $\varphi_{_+}(\omega)$ is still rather
limited. However, the evolution effects shows that for sufficiently
large values of $\mu$, the DA $\varphi_{_{+}}(\omega)$ satisfies the
condition $\varphi_{_+}(\omega) \sim \omega$ as $\omega \rightarrow
0$ and falls off slower than $1/\omega$ for $\omega \rightarrow
\infty$, which implies that the normalization integral of the
$\varphi_{_{+}}$ is ultraviolet divergent. Without considering the
radiative $\mathcal{O}(\alpha_s)$ corrections, the ultraviolet
behavior of the $\varphi_{_{+}}$ plays no role at the leading order
(LO) \cite{BOLange}. Also, the next-to-leading order (NLO) effects
have already been taken into account in more elaborated models of
$\varphi_{_+}$ based on the HQET sum rules \cite{Braun}. In this
work, we use three phenomenological models for the shape of the DA
$\varphi_{_+}$ as \cite{BeneBraun}:
\begin{eqnarray}\label{eq24}
\begin{aligned}
\mbox{Model I}:  \varphi {_{_+}}(\omega)
&=\left[(1-a)+\frac{a\,\omega}{2\,\omega_0}\right]\frac{\omega}{\omega_0^2}\,\,e^{-{\omega}/{\omega_0}},
& 0\leq a \leq 1
&\\
\mbox{Model II}:  \varphi
{_{_+}}(\omega)&=\frac{1}{\Gamma(2+b)}\,\frac{\omega^{1+b}}{\omega_0^{2+b}}
\,e^{-{\omega}/{\omega_0}}, & -0.5 < b < 1
& \\
\mbox{Model III}: \varphi
{_{_+}}(\omega)&=\frac{\sqrt{\pi}}{2\,\Gamma(3/2+c)}\,\frac{\omega}{\omega_0^{2}}
\,e^{-{\omega}/{\omega_0}}\,U(c, 3/2-c, \omega/\omega_0), & 0 < c <
0.5&
\end{aligned}
\end{eqnarray}
where $U(\alpha, \beta, z)$ is the confluent hypergeometric function
of the second kind. In our calculations, we take the upper limiting
values for two parameters $a$ and $c$, hence $a=1$, $c=0.5$. It is
remarkable that for $b=1$, the shape of $\varphi {_{_+}}$ in
model-II become the same as that in model-I for $a=1$, therefore we
take $b=0.5$~. The corresponding expression of
$\varphi_{_-}(\omega)$ for each model is determined by the
equation-of-motion constraint in the absence of contributions from
the three-particle DA's as \cite{BenFel}:
\begin{eqnarray}\label{eq25}
\varphi_{_-}(\omega)=\int_0^1\,\frac{d\tau}{\tau}\,\varphi_{_+}(\omega/\tau).
\end{eqnarray}
The shape parameter $\omega_0$, that is a parameter of $B_s$-meson,
can be converted to $\lambda_{B}(\mu =1~ \mbox{GeV})$ that is the
inverse moment of $\varphi_{_+}(\omega, \mu)$ \cite{BOLange}.
Prediction of the $\lambda_{B}$ value is varied in different models,
for example $\lambda_{B}=(460 \pm 110)~\mbox{MeV}$ calculated using
the two-point QCD sum rules \cite{Braun}, $\lambda_{B} = (460 \pm
160)~ \mbox{MeV}$ estimated via the LCSR approach \cite{Offen},
$\lambda_{B} = (350 \pm 150)~ \mbox{MeV}$ adopted in the QCD
factorization approach \cite{Beneke}, and $\lambda_{B} = (360\pm
110)\,~ \mbox{MeV}$ inferred from analyzing the $\bar{B}_{u} \to
\gamma {l}^{-} \bar{\nu}$ decay by the LCSR \cite{Janowski}. In
addition, a central value $\lambda_{B} > 238~ \rm{MeV}$ has been
provided by the BELLE collaboration at $90\%$ credibility level
\cite{Heller}. The values of $\lambda_{B}$ discussed here, are valid
just for $B^{\pm}$-meson and are applicable for $B_s$ only in the
$\rm{SU(3)}_F$ symmetry limit. Recently, the inverse moment of the
$B_s$-meson distribution amplitude has been predicted from the QCD
sum rules (QCDSR) as $\lambda_{B_s}=(438 \pm 150)~\rm{MeV}$
\cite{KhMaMa}. This value is a reasonable choice for the numerical
analysis of the semileptonic $B_s\to K_0^*$ form factors. In this
work, we take $\omega_0=\lambda_{B_s}$ and use the value $(438 \pm
150)~\rm{MeV}$ for it. The dependence of the two-particle DA's with
respect to $\omega$ is shown in Fig. \ref{F2} for the three models
in Eq. (\ref{eq24}).

Comparing to the two-particle DA's, the contribution of the
three-particle DA's is less than $10\%$ in calculations of the form
factors. The three-particle DA's are related to a basis of DA's such
as $\phi_{3}, \phi_{4}, \psi_{4}$ and $\psi_{5}$ with definite
twist, as follows \cite{JiMan}:
\begin{eqnarray}\label{eq26}
\begin{aligned}
\Psi_{_A}(\omega, \xi)      & = \frac{1}{2}\, [ \phi_{3}(\omega,
\xi) + \phi_{4}(\omega, \xi)]\,, &
\Psi_{_V}(\omega, \xi)      & = \frac{1}{2}\, [-\phi_{3}(\omega, \xi) + \phi_{4}(\omega, \xi)]\,, \\
X_{_A}(\omega, \xi)         & = \frac{1}{2}\, [-\phi_{3}(\omega,
\xi) - \phi_{4}(\omega, \xi) + 2\,\psi_{4}(\omega, \xi)]\,,&
Y_{_A}(\omega, \xi)         & = \frac{1}{2}\, [-\phi_{3}(\omega,
\xi) - \phi_{4} (\omega, \xi)+  \psi_{4}(\omega, \xi) -
\psi_{5}(\omega, \xi)]\,.
\end{aligned}
\end{eqnarray}
So far, several models have been proposed for the shape of
$\phi_{3}, \phi_{4}, \psi_{4}$ and $\psi_{5}$. Since the structure
of $\varphi_{_+}$ in three models in Eq. (\ref{eq24}) is the
exponential form, we choose the exponential model for the functions
$\phi_{3}, \phi_{4}, \psi_{4}$ and $\psi_{5}$, presented as
\cite{JiMan,ShenWangWei}:
\begin{eqnarray}\label{eq26}
\begin{aligned}
\phi_{3}(\omega, \xi) &= {\lambda_E^2 - \lambda_H^2 \over 6 \,
\omega_0^5} \, \omega \, \xi^2 \, e^{-\frac{\omega + \xi}{
\omega_0}}\,, &
\phi_{4}(\omega, \xi) &= {\lambda_E^2 + \lambda_H^2
\over 6 \, \omega_0^4} \, \xi^2 \,
e^{-\frac{\omega + \xi}{ \omega_0}} \,, \\
\psi_{4}(\omega, \xi) &= {\lambda_E^2 \over 3 \, \omega_0^4} \,
\omega \, \xi \, e^{-\frac{\omega + \xi}{ \omega_0}}\,,&
\psi_{5}(\omega, \xi) &= - {\lambda_E^2 \over 3 \, \omega_0^3} \,
\xi \, e^{-\frac{\omega + \xi}{ \omega_0}} \,.
\end{aligned}
\end{eqnarray}

\begin{figure}[th]
\begin{center}
\includegraphics[width=5cm,height=5cm]{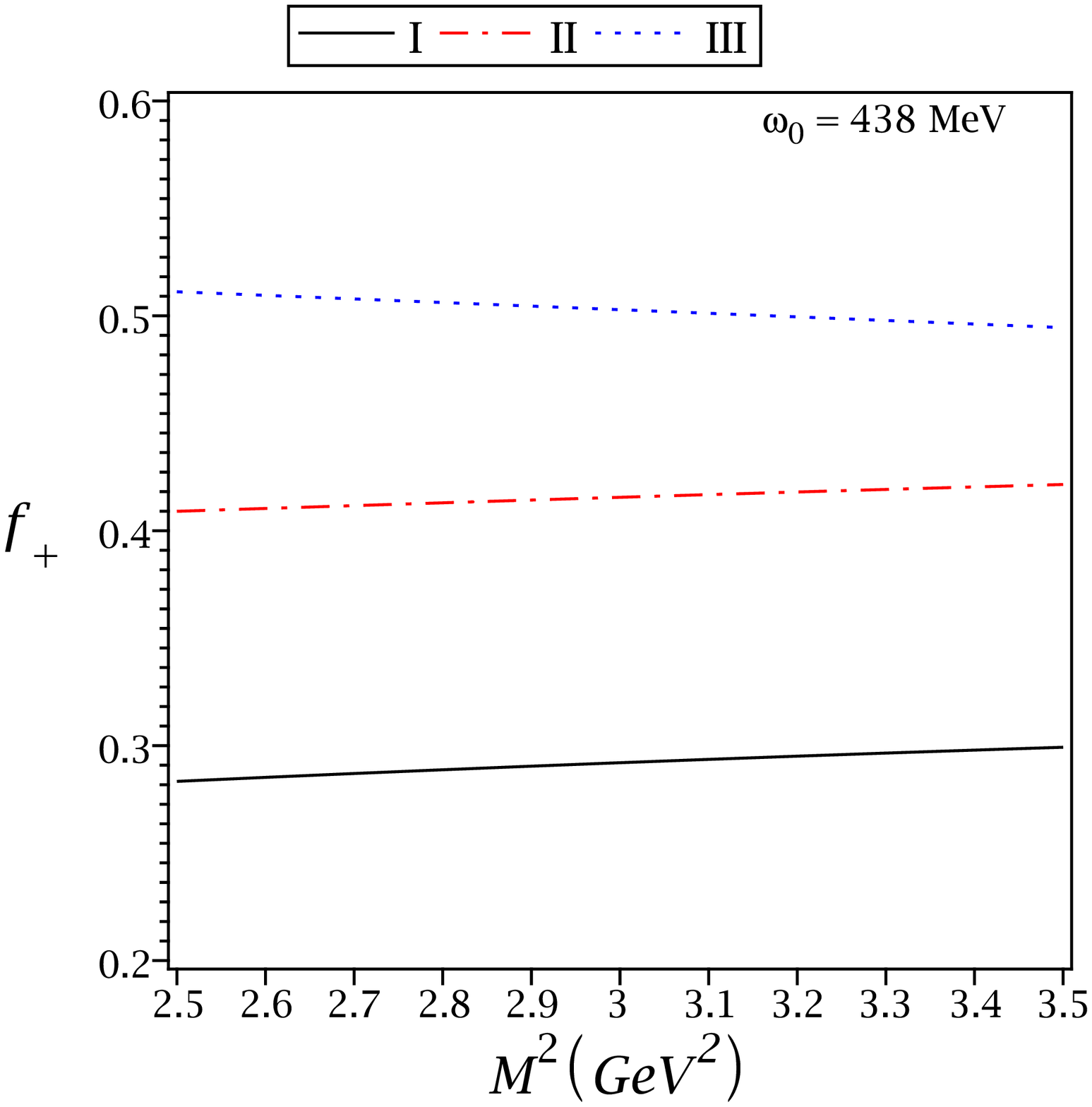}
\includegraphics[width=5cm,height=5cm]{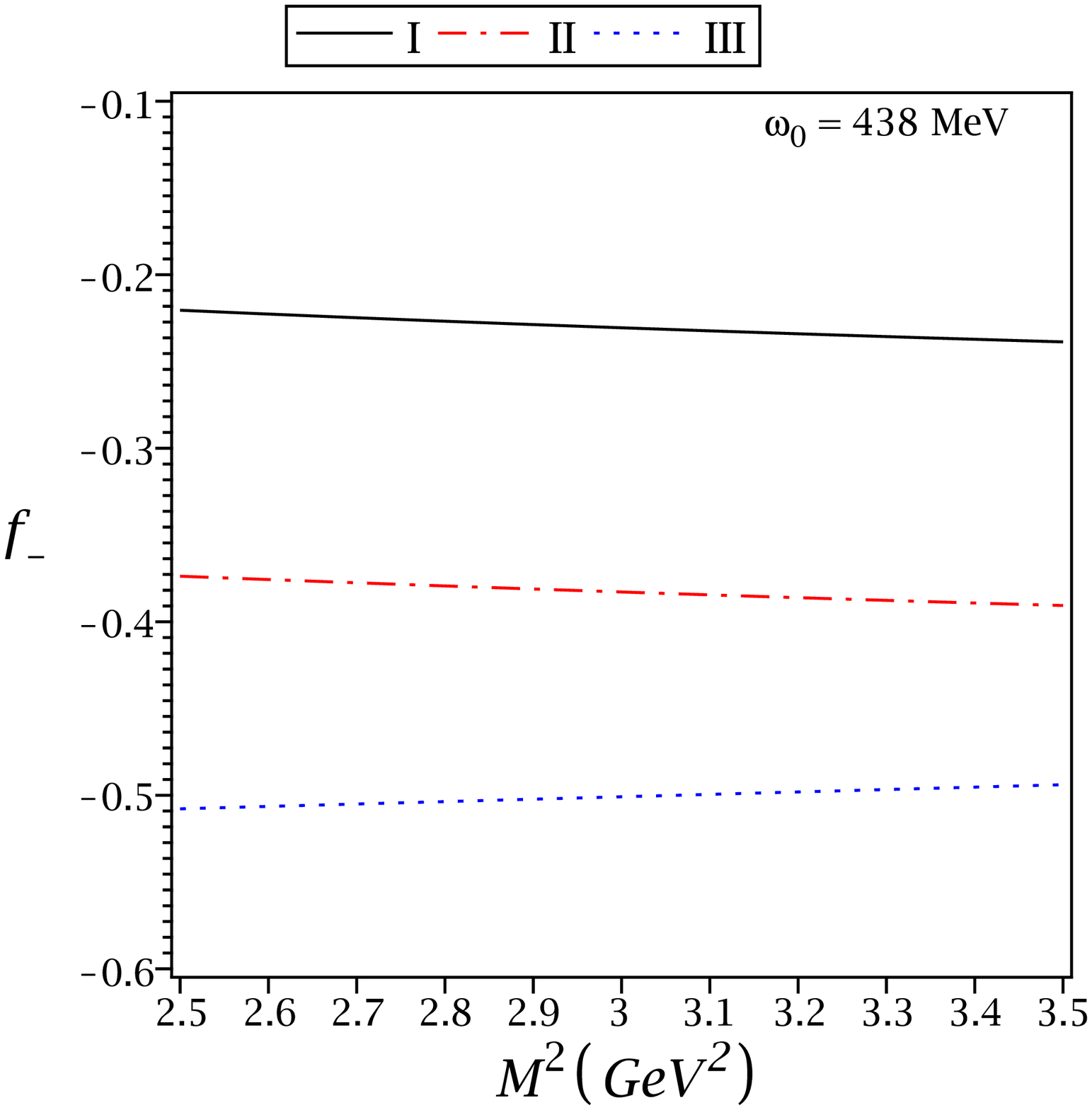}
\includegraphics[width=5cm,height=5cm]{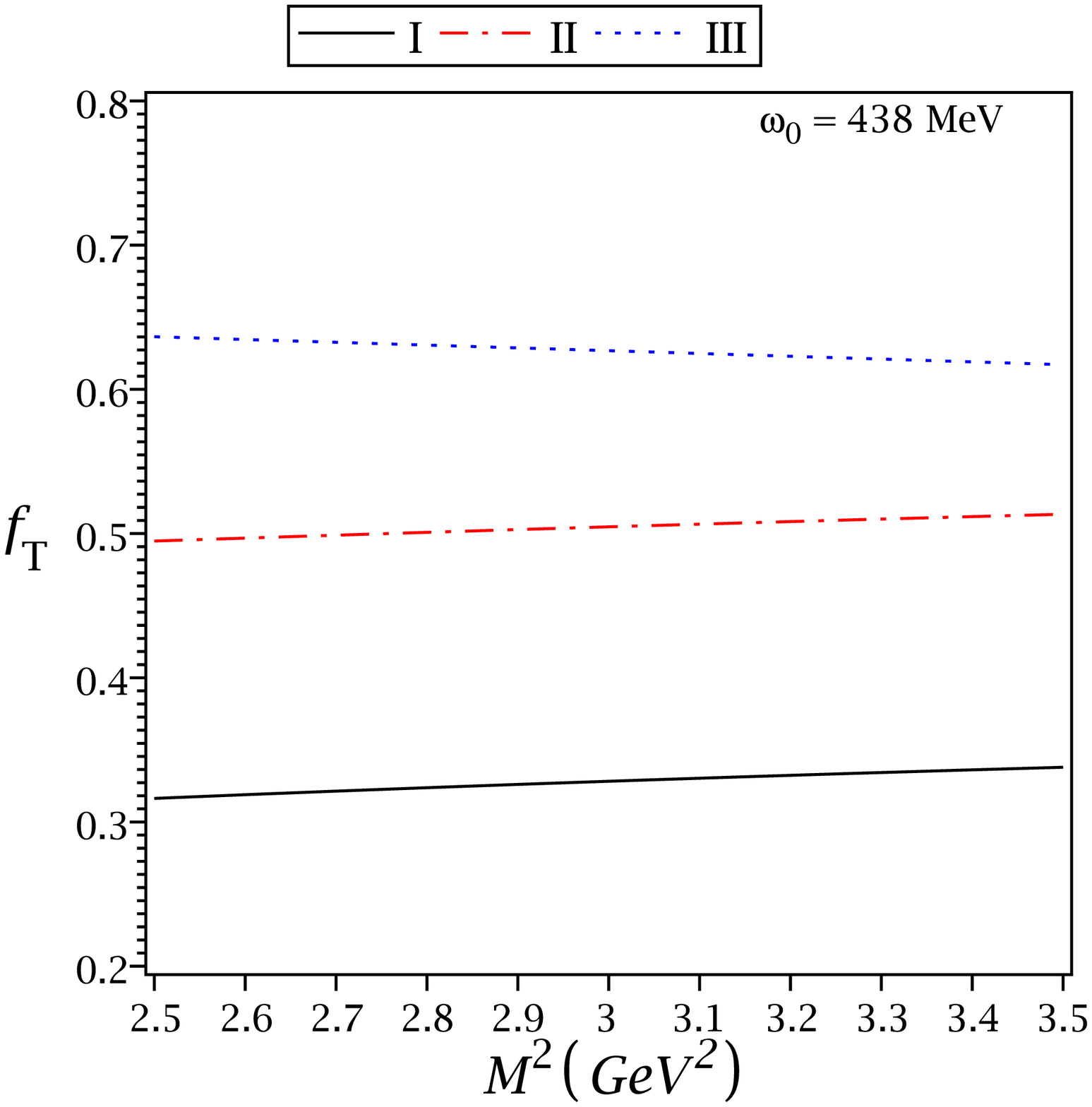}
\caption{The dependence of the form factors $f_+$, $f_{-}$ and
$f_{T}$ on the Borel parameter $M^2$ for the three models in
$\omega_0=438\, \rm{MeV}$, and $q^2=0\, \rm{GeV^2}$.} \label{F3}
\end{center}
\end{figure}

To analyze the form factors $f_{+}$, $f_{-}$, and $f_{T}$, the value
of the Borel parameter $M^2$ must also be determined. The Borel
parameter $M^2$  is not physical quantity, so the physical
quantities, form factors, should be independent of it. The working
region for $M^2$ is determined by requiring that the contributions
of the higher states and continuum are effectively suppressed. The
dependence of the form factors $f_+$, $f_{-}$ and $f_{T}$ on the
Borel parameter $M^2$ is shown in Fig. \ref{F3}, for the three
models in $\omega_0=438\, \rm{MeV}$, and $q^2=0\, \rm{GeV^2}$.
This figure shows a good stability of the form factors with respect
to the Borel parameter in the  interval: $2.5\, \mbox{GeV}^2\leq
M^2\leq 3.5\,\mbox{GeV}^2$. We take $M^2=3~ \mbox{GeV}^2$ in our
calculations. Uncertainties originated from the Borel parameter
$M^2$ in this  region, are about $5\%$.

Having all these input values and parameters, we proceed to carry
out numerical calculations. Inserting the values of the masses,
leptonic decay constants, continuum threshold, Borel mass, the
parameters of the $B_s$-meson DA's such as $\omega_0$ and other
quantities that appear in the form factors in Eq. (\ref{eq22}), we
can calculate the form factors of the semileptonic $B_s\to K_0^*$
transitions at zero momentum transfer, $q^2=0\,\mbox{GeV}^2$. Table
\ref{T1} shows central values of the form factors for the three
models as well as sources of error and also uncertainties caused by
them, separately. As can be seen $\omega_{0}$ and $f_{K_0^*}$ are
the most significant sources of theory uncertainties.
\begin{table}[th]
\caption{Central values of the form factors for the three models, as
well as sources of error and also uncertainties of the form factors.
The uncertainties $\Delta$ caused by the variations of the input
parameters ($\delta {\omega_{0}}=\pm 0.150~\mbox{GeV}$,  $\delta
{f_{K_0^*}}=\pm 0.085 ~\mbox{GeV}$,  $\delta {s_{0}}=\pm
0.4~\mbox{GeV}^2$, $\delta {m_{K_{0}^{*}}}=\pm 0.05~\mbox{GeV}$,
$\delta {f_{B_s}}=\pm 0.001~\mbox{GeV}$, $\delta
{\lambda_{E}^{2}}=\pm 0.01~\mbox{GeV}^2$, $\delta
{\lambda_{H}^{2}}=\pm 0.05~\mbox{GeV}^2$, $\delta { M^2}=\pm
0.5~\mbox{GeV}^2$).}\label{T1}
\begin{ruledtabular}
\begin{tabular}{ccccccccccc}
\rm{Model} & \rm{Form Factor} & \rm{Central Value} &
$\Delta(\omega_{0})$ &$\Delta(f_{K_0^*})$ & $\Delta( s_{0})$ &
$\Delta( m_{K_{0}^{*}})$ & $\Delta( f_{B_s})$ & $\Delta(
\lambda_{E}^{2})$ & $\Delta( \lambda_{H}^{2})$ & $\Delta( M^2)$
\\
\hline\\[-2mm]
           &$f_{+}(0)$      & $+0.283$                 & $^{+0.252}_{-0.114} $ & $^{+0.070}_{-0.047} $ & $^{+0.024}_{-0.027} $ & $^{+0.004}_{-0.003} $ & $^{+0.002}_{-0.003} $ & $^{+0.000}_{-0.000} $ & $^{+0.004}_{-0.004} $ & $^{+0.006}_{-0.007} $\\[2mm]

\rm{I}     &$f_{-}(0)$      & $-0.228$                 & $^{+0.114}_{-0.262} $ & $^{+0.038}_{-0.056} $ & $^{+0.027}_{-0.024} $ & $^{+0.003}_{-0.003} $ & $^{+0.002}_{-0.002} $ & $^{+0.002}_{-0.002} $ & $^{+0.002}_{-0.001} $ & $^{+0.011}_{-0.008} $\\[2mm]

           &$f_{T}(0)$      & $+0.324$                 & $^{+0.326}_{-0.145} $ & $^{+0.080}_{-0.054} $ & $^{+0.030}_{-0.034} $ & $^{+0.007}_{-0.006} $ & $^{+0.003}_{-0.003} $ & $^{+0.000}_{-0.001} $ & $^{+0.001}_{-0.002} $ & $^{+0.009}_{-0.012} $\\[3mm]
\hline&&&&&&&& \\[-2mm]
           &$f_{+}(0)$      & $+0.412$                 & $^{+0.279}_{-0.145} $ & $^{+0.102}_{-0.069} $ & $^{+0.025}_{-0.030} $ & $^{+0.006}_{-0.005} $ & $^{+0.003}_{-0.004} $ & $^{+0.000}_{-0.000} $ & $^{+0.003}_{-0.004} $ & $^{+0.005}_{-0.005} $\\[2mm]

\rm{II}    &$f_{-}(0)$      & $-0.369$                 & $^{+0.149}_{-0.293} $ & $^{+0.061}_{-0.092} $ & $^{+0.029}_{-0.027} $ & $^{+0.004}_{-0.006} $ & $^{+0.003}_{-0.004} $ & $^{+0.001}_{-0.001} $ & $^{+0.001}_{-0.002} $ & $^{+0.009}_{-0.009} $\\[2mm]

           &$f_{T}(0)$      & $+0.495$                 & $^{+0.363}_{-0.186} $ & $^{+0.123}_{-0.082} $ & $^{+0.033}_{-0.038} $ & $^{+0.011}_{-0.009} $ & $^{+0.004}_{-0.004} $ & $^{+0.001}_{-0.001} $ & $^{+0.002}_{-0.001} $ & $^{+0.009}_{-0.009} $\\[3mm]
\hline&&&&&&&& \\[-2mm]
           &$f_{+}(0)$      & $+0.511$                 & $^{+0.195}_{-0.124} $ & $^{+0.127}_{-0.085} $ & $^{+0.015}_{-0.018} $ & $^{+0.008}_{-0.005} $ & $^{+0.005}_{-0.004} $ & $^{+0.000}_{-0.000} $ & $^{+0.004}_{-0.003} $ & $^{+0.013}_{-0.007} $\\[2mm]

\rm{III}   &$f_{-}(0)$      & $-0.506$                 & $^{+0.127}_{-0.188} $ & $^{+0.084}_{-0.125} $ & $^{+0.019}_{-0.014} $ & $^{+0.006}_{-0.007} $ & $^{+0.005}_{-0.004} $ & $^{+0.001}_{-0.001} $ & $^{+0.002}_{-0.001} $ &  $^{+0.005}_{-0.008} $\\[2mm]

           &$f_{T}(0)$      & $+0.644$                 & $^{+0.243}_{-0.158} $ & $^{+0.161}_{-0.107} $ & $^{+0.019}_{-0.023} $ & $^{+0.014}_{-0.011} $ & $^{+0.006}_{-0.005} $ & $^{+0.001}_{-0.000} $ & $^{+0.002}_{-0.001} $ & $^{+0.013}_{-0.007} $\\[2mm]
\end{tabular}
\end{ruledtabular}
\end{table}

Taking into account all the uncertainty values except $\omega_0$,
the numerical values of the form factors $f_{+}$, $f_{-}$ and
$f_{T}$ in $q^2=0\,\mbox{GeV}^2$ are presented in Table \ref{T3} for
the three models. This table also includes a comparison of our
results with the predictions of other approaches such as the LCSR
with the light-meson DA's \cite{HanWuFu,YMWang,YJSun}, perturbative
QCD (PQCD) \cite{RHLi} and QCDSR method \cite{MZYang,Ghahramany}. As
can be seen, there is a very good agreement between our results in
model II and predictions of the conventional LCSR with the
light-meson DA's in S2 \cite{YMWang}. As a result, our calculations
confirm scenario 2 for describing the scalar meson $K_0^*$.
\begin{table}[th]
\caption{The form factors of the semileptonic $B_{s} \to K_0^*$
transitions at zero momentum transfer from the three models and
different approaches.}\label{T3}
\begin{ruledtabular}
\begin{tabular}{ccccc}
\rm{Method}    &  $f_{+}(0)$ & $f_{-}(0)$ & $f_{T}(0)$ \\
\hline\\[-2mm]
\quad\quad\quad\quad\quad\,(I)    & ${+0.28}^{+0.11}_{-0.09}$ & ${-0.10}^{+0.09}_{-0.19}$ & ${+0.32}^{+0.13}_{-0.11}$\\[2mm]
{This work} \,\,\,{(II)}          & ${+0.41}^{+0.14}_{-0.12}$ & ${-0.37}^{+0.11}_{-0.14}$ & ${+0.50}^{+0.18}_{-0.14}$\\[2mm]
\quad\quad\quad\quad\quad\,(III)    & ${+0.51}^{+0.16}_{-0.12}$ & ${-0.51}^{+0.12}_{-0.16}$ & ${+0.64}^{+0.22}_{-0.15}$\\[2mm]
LCSR(S2) \,\,\,\cite{YMWang}      & $+0.42^{+0.13}_{-0.08}  $ & $-0.34^{+0.10}_{-0.10}  $ & $+0.52^{+0.18}_{-0.08}$ \\[2mm]
LCSR(S2) \,\,\,\cite{HanWuFu}     & $+0.39^{+0.04}_{-0.04}  $ & $-0.25^{+0.05}_{-0.05}  $ & $+0.41^{+0.04}_{-0.04}$ \\[2mm]
LCSR(S2) \,\,\,\cite{YJSun}       & $+0.44                  $ & $-0.44                  $ & $--                 $ \\[2mm]
PQCD(S2)\,\,\,\cite{RHLi}         & $+0.56^{+0.16}_{-0.13}  $ & $--$ & $+0.72^{+0.27}_{-0.17} $ \\[2mm]
LCSR(S1) \,\,\,\cite{YJSun}       & $+0.10                  $ & $-0.10                  $ & $--                 $ \\[2mm]
PQCD(S1)\,\,\,\cite{RHLi}         & $-0.32^{+0.06}_{-0.07}  $ & $--$ & $-0.41^{+0.08}_{-0.09} $ \\[2mm]
QCDSR \,\,\,\,\,\,\,\cite{MZYang}    & $+0.24 \pm 0.10         $ & $--                   $ & $--            $ \\[2mm]
QCDSR \,\,\,\,\,\,\,\cite{Ghahramany} & $+0.25 \pm 0.05         $ & $-0.17 \pm 0.04       $ & $+0.21 \pm 0.04 $ \\[2mm]
\end{tabular}
\end{ruledtabular}
\end{table}

Due to the presence of cutoff in the QCD calculations, we look for a
parametrization of the form factors to extend our results to the
full physical region, $0 \leq q^2 \leq {(m_{B_s} -
m_{K_{0}^{*}})}^2$. Through fitting the results of the LCSR among
the region $0 < q^2 < 8\, \rm {GeV}^2$, we extrapolate them with the
pole model parametrization
\begin{eqnarray}\label{eq27}
f_{i}(q^{2})=\frac{f_i(0)}{1-\alpha(q^2/m_{B_{s}}^2)+\beta
{(q^2/m_{B_{s}}^2)}^2},
\end{eqnarray}
with the constants $\alpha$ and $\beta$ determined from the fitting
procedure. The values of the parameters $\alpha$ and $\beta$ are
presented in Table \ref{T4} for the three models. The values of
parameter $f_{i}(0)$ expressed the form factor results at
$q^2=0\,\mbox{GeV}^2$ were listed in Table \ref{T3}, before.
\begin{table}[th]
\caption{The parameters $\alpha$ and $\beta$ obtained for the form
factors of the semileptonic $B\to K_0^*$ transitions for the three
models.} \label{T4}
\begin{ruledtabular}
\begin{tabular}{cccccccccccc}
\rm{Form
Factor}&\multicolumn{3}{c}{$f_{+}(q^2)$}&&\multicolumn{3}{c}{$f_{-}(q^2)$}&&\multicolumn{3}{c}{$f_{T}(q^2)$}\\
\cline{2-4}\cline{6-8}\cline{10-12}\rm{Model}&I&II&III&&I&II&III&&I&II&III
\\ \hline
$\alpha$         & $-0.14$&$-0.24$ &$-0.49$ & & $+0.09$&$+0.19$ &$+0.46$ & & $-0.14$&$-0.27$ &$-0.60$\\
$\beta $& $+0.26$&$+0.63$ &$+3.89$ && $-0.14$&$-0.45$&$-3.76$ &&
$+0.25$&$+0.69$ &$+5.00$
\end{tabular}
\end{ruledtabular}
\end{table}

\begin{figure}[th]
\begin{center}
\includegraphics[width=5cm,height=5cm]{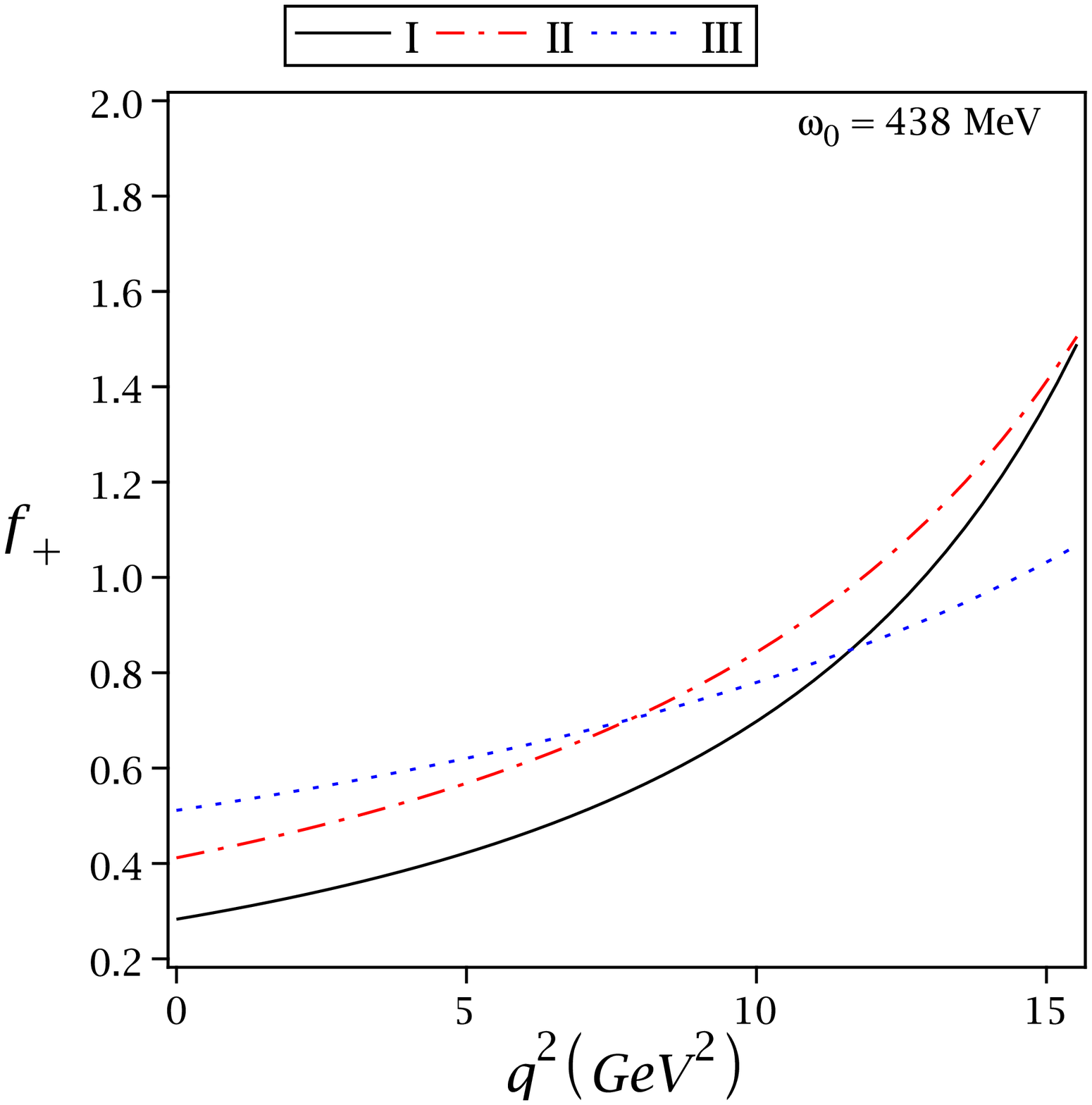}
\includegraphics[width=5cm,height=5cm]{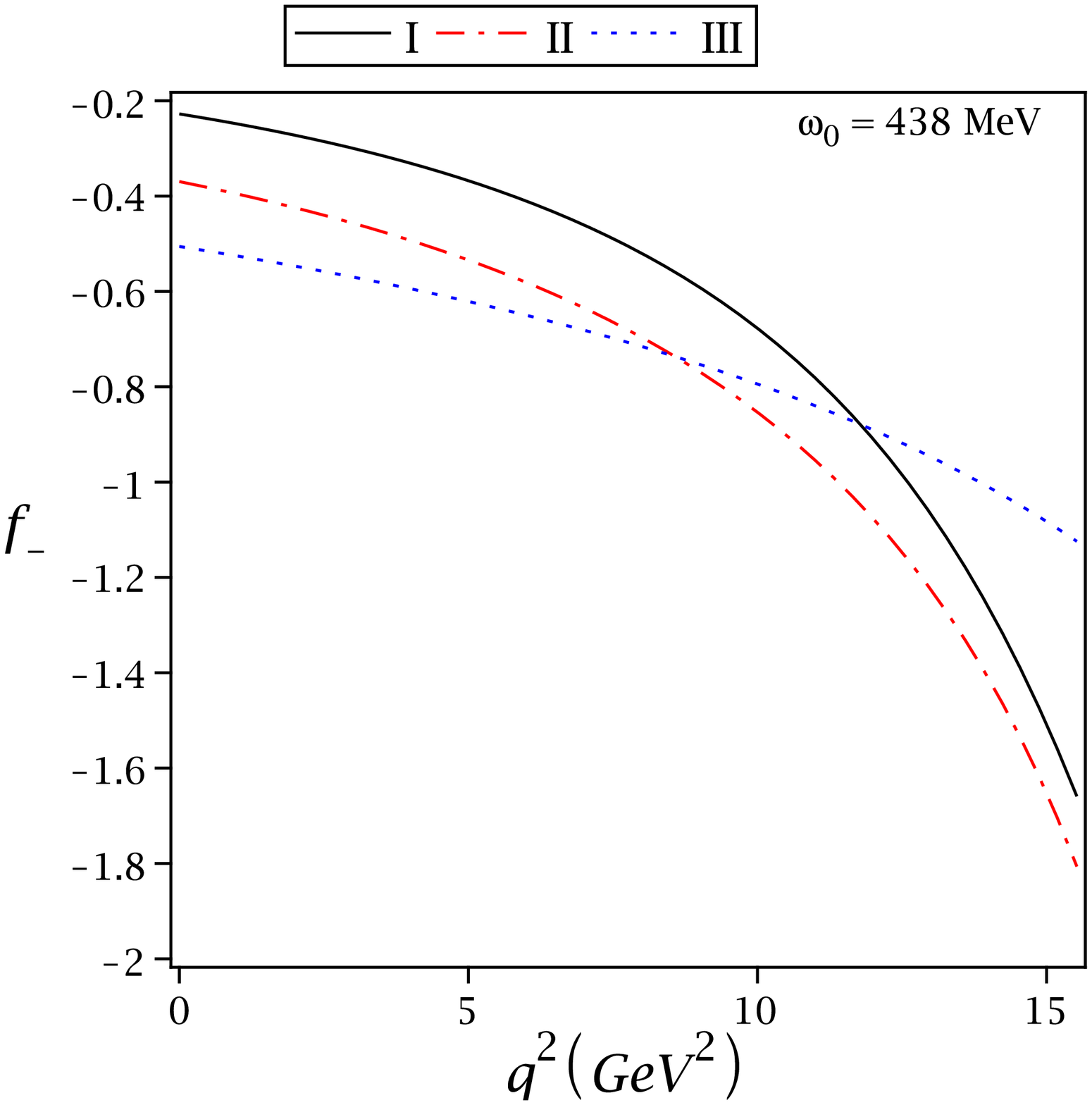}
\includegraphics[width=5cm,height=5cm]{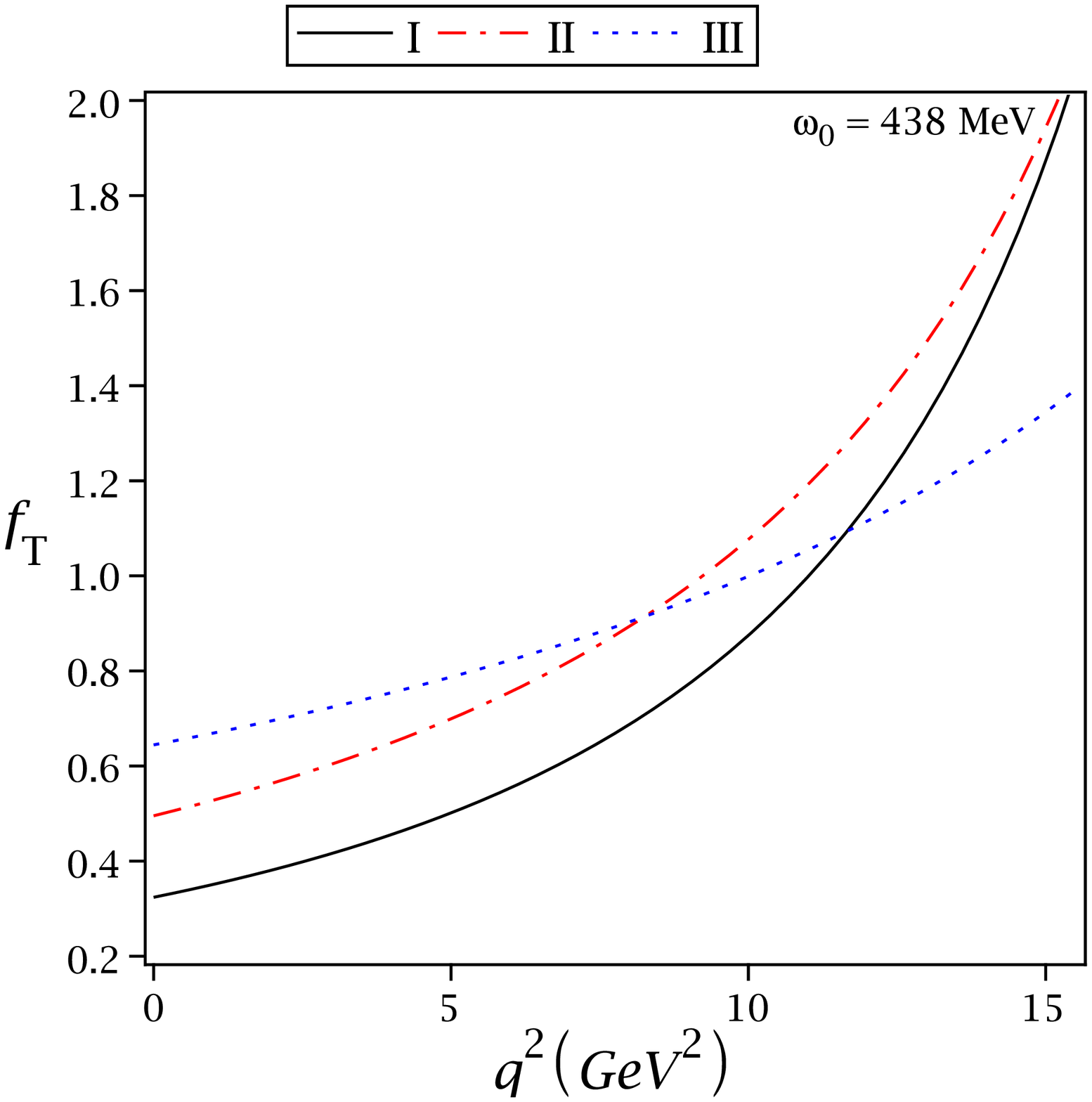}
\caption{The dependence of the form factors $f_{+}(q^2)$,
$f_{-}(q^2)$ and $f_{T}(q^2)$ of the semileptonic $B_s \rightarrow
K_0^* $ transitions on $q^2$ for the three models.} \label{F4}
\end{center}
\end{figure}

\begin{figure}[th]
\begin{center}
\includegraphics[width=5cm,height=5cm]{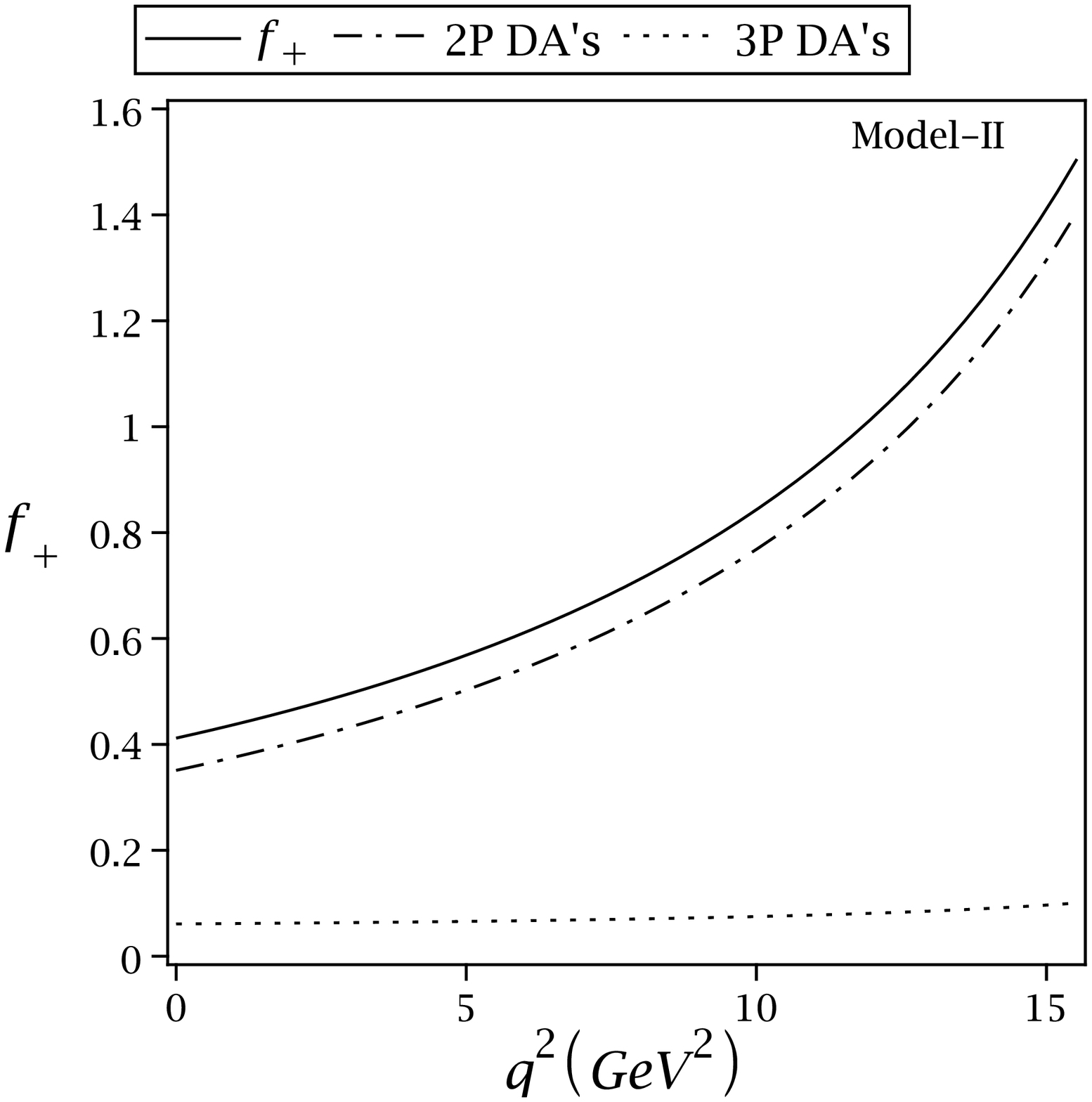}
\includegraphics[width=5cm,height=5cm]{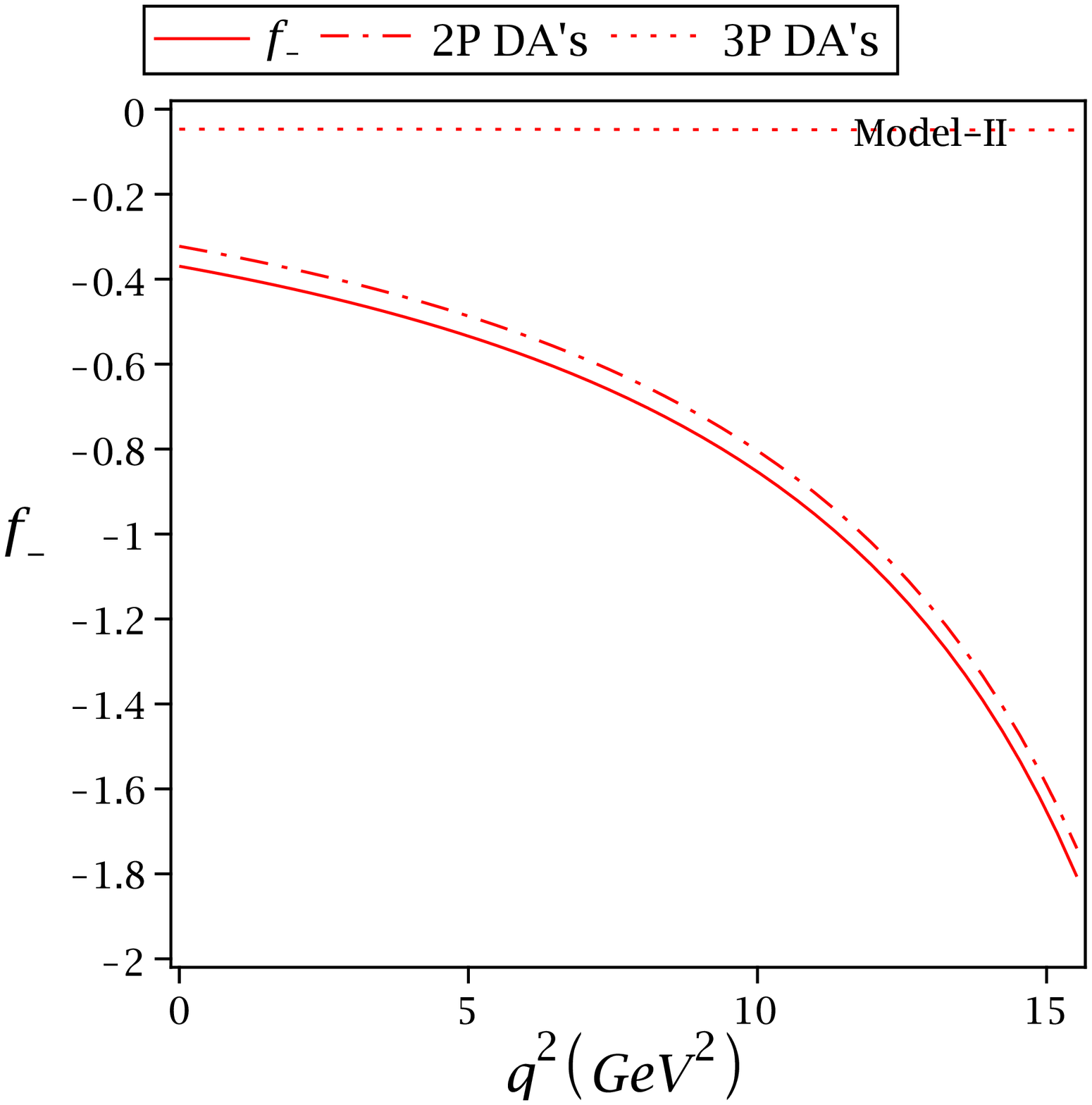}
\includegraphics[width=5cm,height=5cm]{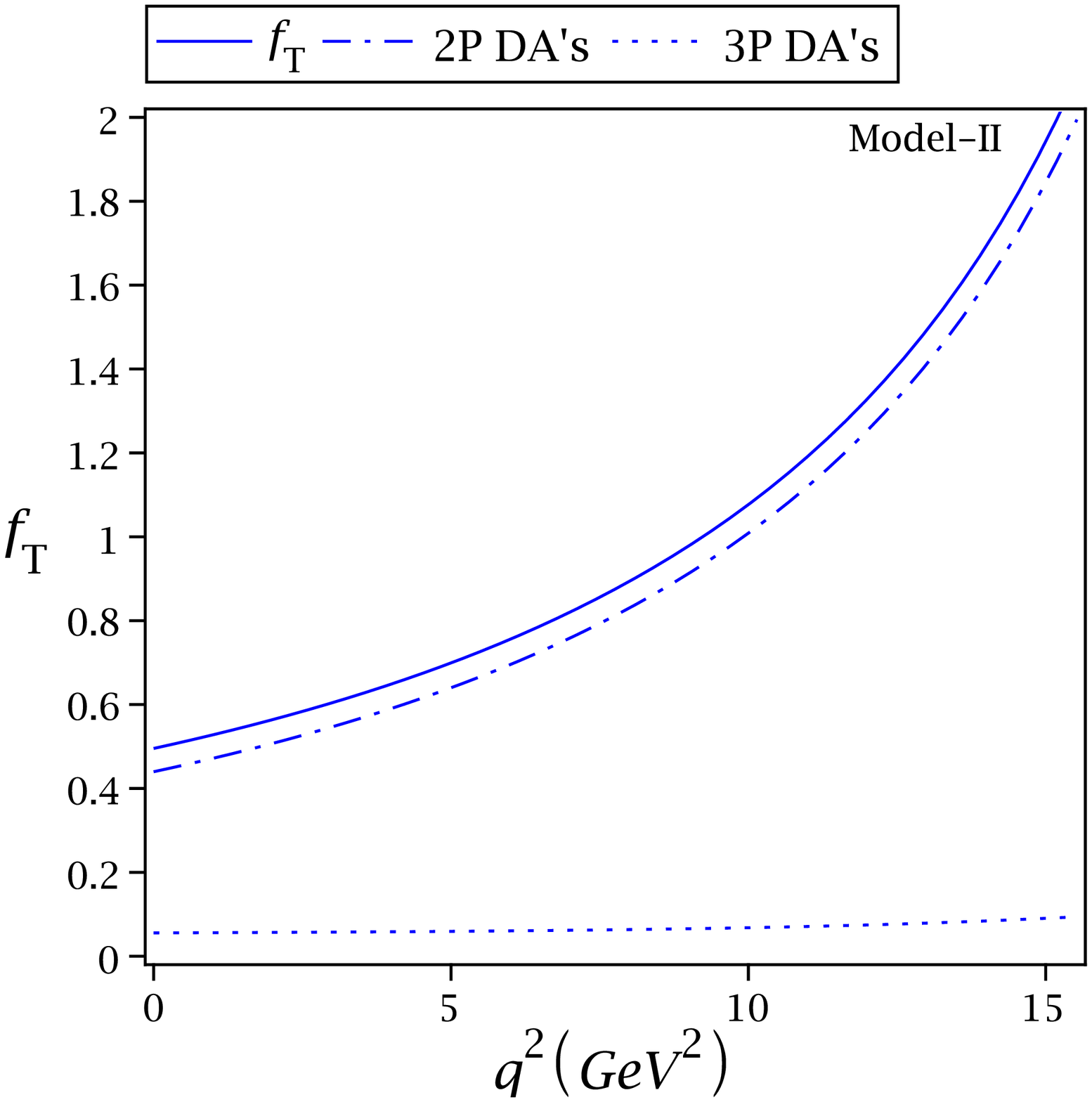}
\caption{The contributions of the two-particle DA's (2P DA's) and
three-particle DA's (3P DA's) in the form factors $f_+(q^2)$,
$f_{-}(q^2)$ and $f_{T}(q^2)$ for model II.} \label{F05}
\end{center}
\end{figure}

The dependence of the form factors $f_{+}$, $f_{-}$ and $f_{T}$ on
$q^2$, for the three models, is shown in Fig. \ref{F4}. In this
work, the form factors are estimated in the LCSR approach up to the
three-particle DA's of the $B_s$-meson.
Our calculations show that the most contributions comes from the
two-particle functions $\varphi_{\pm}$ for all form factors, so that
the contributions of the three-particle DA's are less than $10 \%$
of the total. The contributions of the two- and three-particle DA's
in the form factors depict in Fig. \ref{F05} for model II,
separately.

\begin{figure}[th]
\begin{center}
\includegraphics[width=5cm,height=5cm]{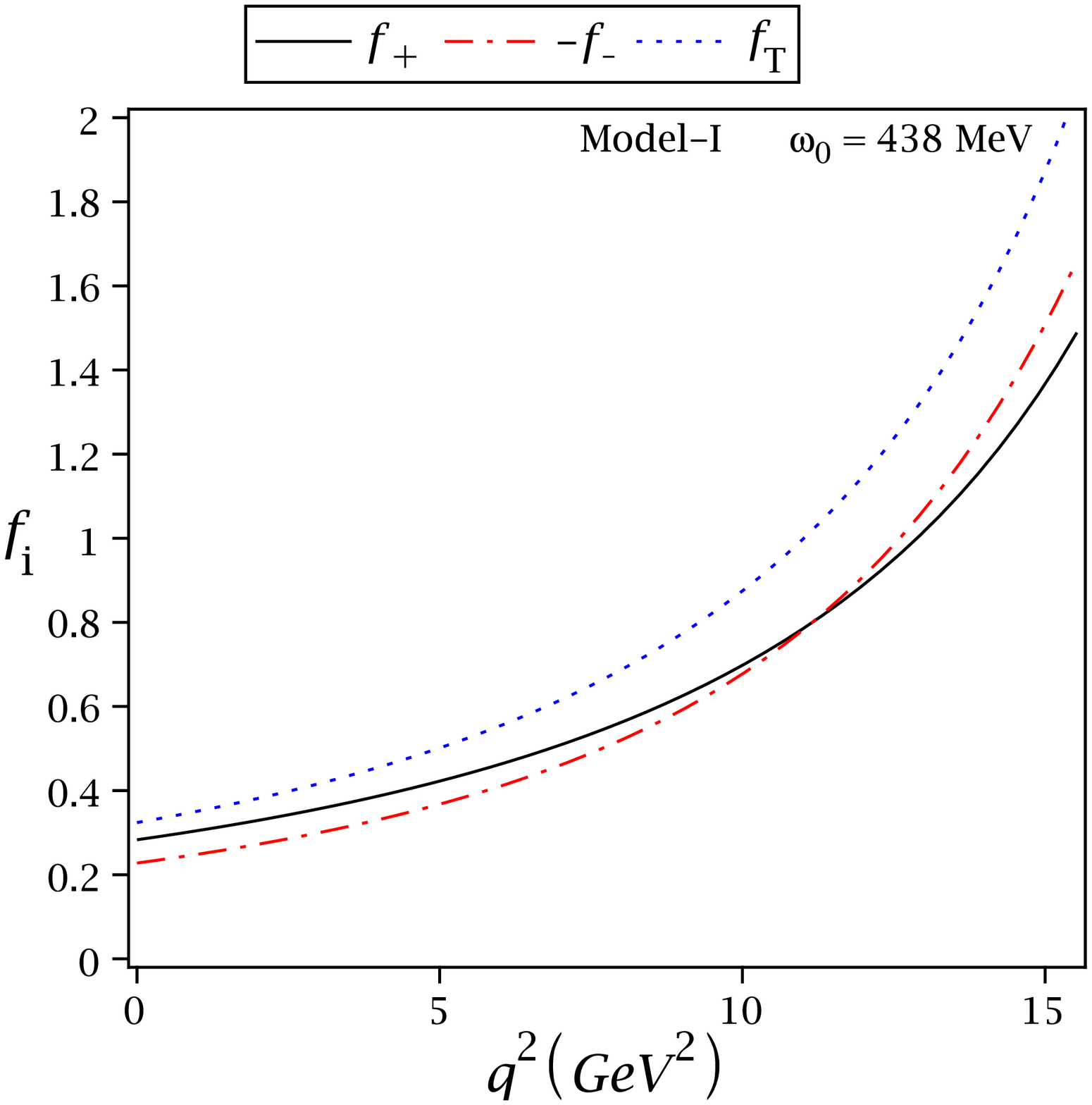}
\includegraphics[width=5cm,height=5cm]{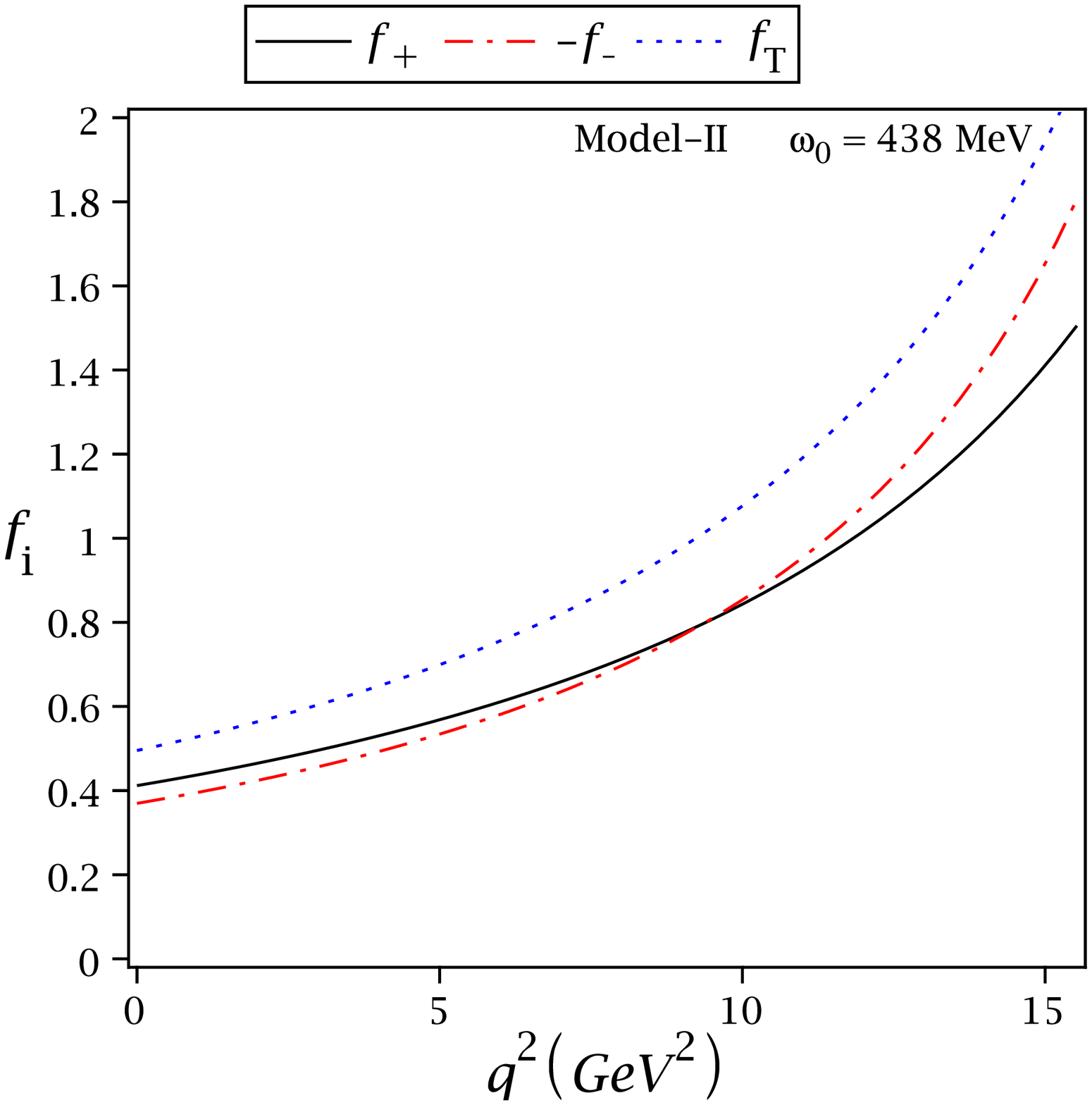}
\includegraphics[width=5cm,height=5cm]{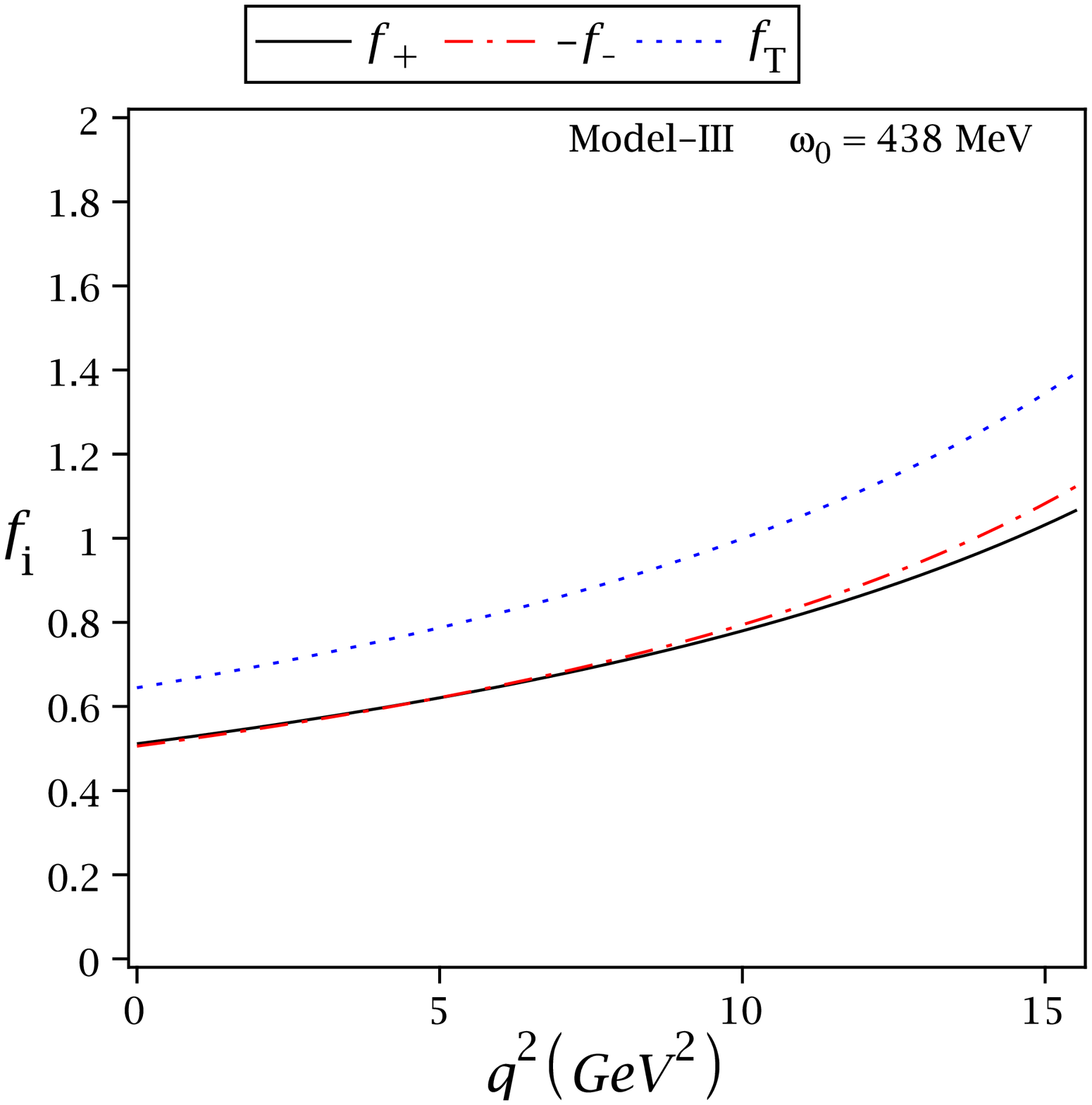}
\caption{The dependence of the form factors $f_+(q^2)$,
$-f_{-}(q^2)$ and $f_{T}(q^2)$ on $q^2$ via the LCSR with the
$B_s$-meson DA's for the three models.} \label{F5}
\end{center}
\end{figure}

The form factors at large recoil should satisfy the following
relations \cite{ColFazWan}:
\begin{eqnarray}\label{eq28}
f_{T}(q^2)=\frac{m_{B_s}+m_{K_0^*}}{m_{B_s}}f_{+}(q^2)=-\frac{m_{b}}{m_{B_s}-m_{K_0^*}}f_{-}(q^2).
\end{eqnarray}
Figuer \ref{F5} shows that the computed form factors from the LCSR
with the $B_s$-meson DA's for the three models satisfy the relations
in Eq. (\ref{eq28}), by considering the errors.

With the derived transition form factors, one can proceed to perform
the calculations on some interesting observables in phenomenology,
such as decay rate, polarization asymmetry, and forward-backward
asymmetry. Note that the forward-backward asymmetry for the decay
mode $B_s \to K_0^* l^+ l^-$ is exactly equal to zero in the SM
\cite{BelGen}.

\begin{figure}[th]
\begin{center}
\includegraphics[width=6cm,height=6cm]{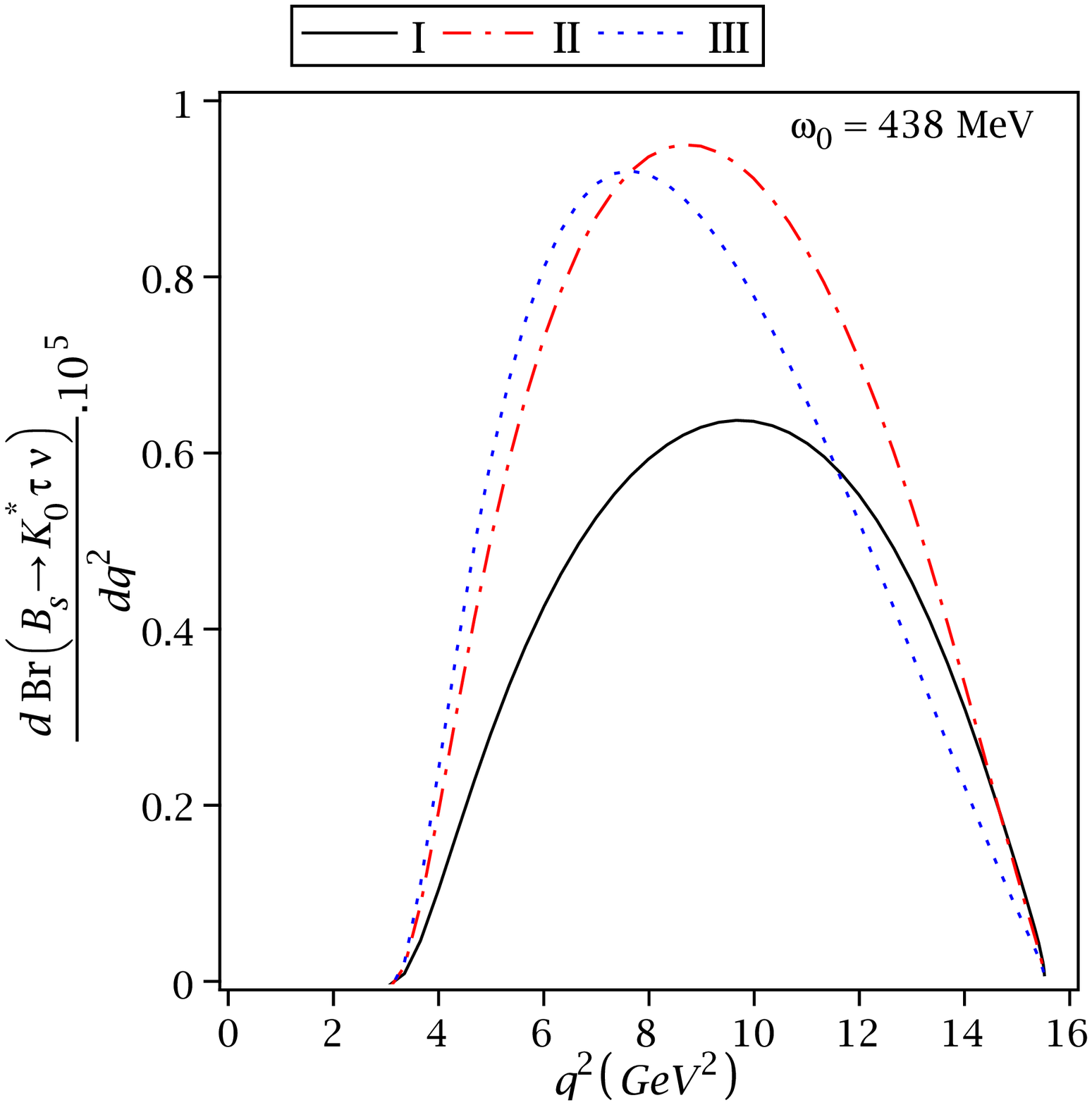}
\includegraphics[width=6cm,height=6cm]{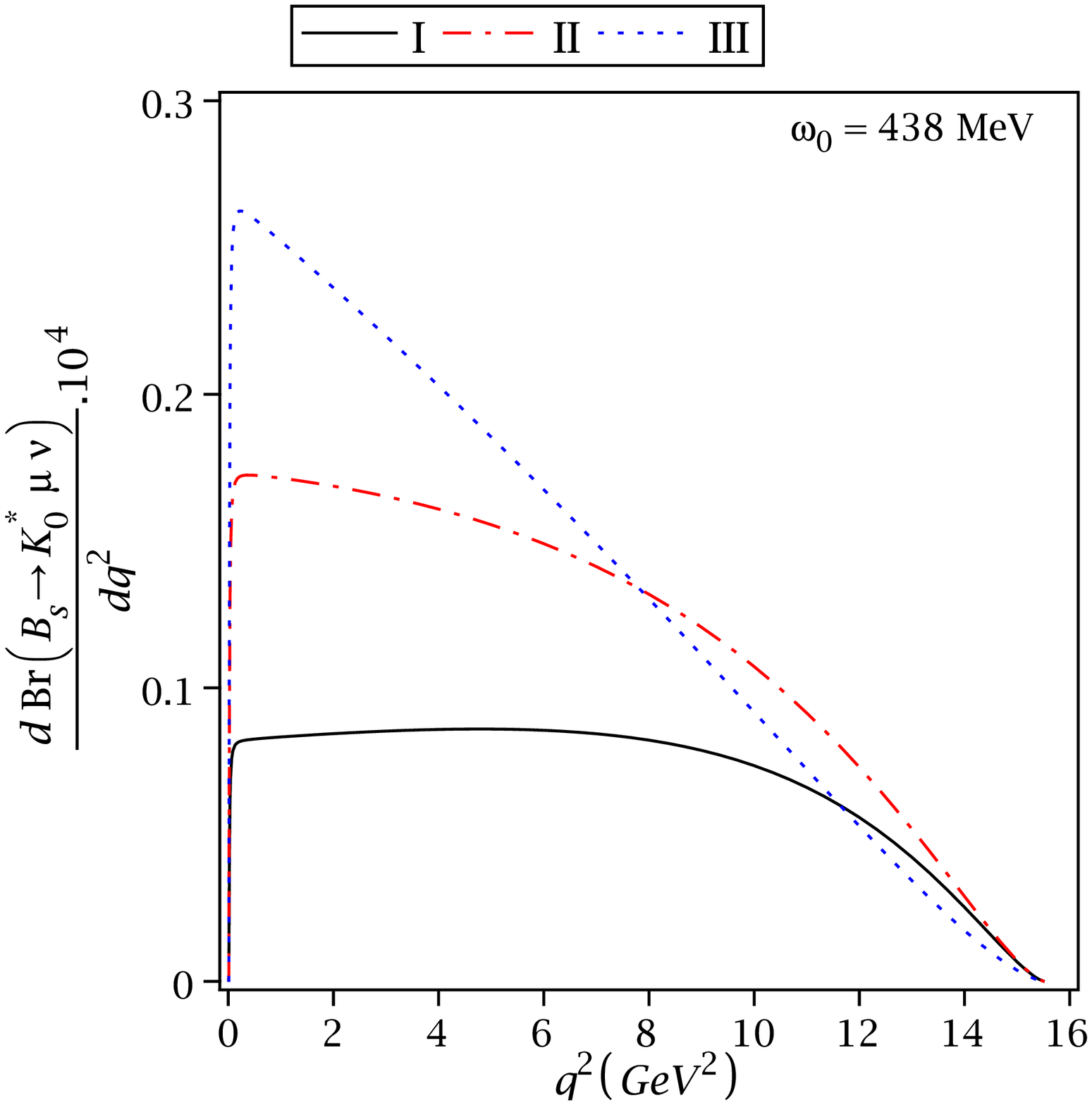}
\caption{Differential branching ratios of the semileptonic $B \to
K_0^* l \nu_{l}$ decays on $q^2$ for the three models.} \label{F6}
\end{center}
\end{figure}

The effective Hamiltonian for $b\to u l \bar{\nu}_{l}$ transition is
\begin{eqnarray}\label{eq29}
\mathcal{H}_{\rm{eff}}(b\to u l
\bar{\nu}_{l})=\frac{G_{F}}{\sqrt{2}} V_{ub}\, \bar{u}\gamma_{\mu}
(1-\gamma_{5}) b\,\, \bar{l} \gamma^{\mu} (1-\gamma_{5}) \nu_{l}\,.
\end{eqnarray}
With this Hamiltonian, the $q^2$ dependant decay width
$\frac{d\Gamma }{dq^2}$ can be expressed as \cite{YJSun}
\begin{eqnarray}\label{eq30}
&&\frac{d\Gamma}{dq^2}(B_{s}\to K_0^*\, l\,
\bar{\nu}_l)=\frac{G_F^2|V_{ub}|^2}{384\, \pi^3\,
m_{B_s}^3}\frac{(q^2-m_l^2)^2}{(q^2)^3} \sqrt{(m_{B_s}^2-m_{
K_0^*}^2-q^2)^2 -4\, q^2\, m_{ K_0^*}^2}
\,\,\,\Big\{(m_l^2+2q^2) \nonumber\\
&& \times \sqrt{(m_{B_s}^2-m_{ K_0^*}^2-q^2)^2 -4 q^2 m_{
K_0^*}^2}\,\,f_+^2(q^2)
+3\,m_l^2\,(m_{B_s}^2-m_{K_0^*}^2)^2\,\Big[f_+(q^2)+\frac{q^2}{m_{B_s}^2-m_{
K_0^*}^2}f_-(q^2)\Big]^2\Big\},
\end{eqnarray}
where $V_{ub}=(3.82\pm0.24)\times 10^{-3}$, and $m_{l}$ is the mass
of the lepton. Integrating Eq. (\ref{eq30}) over $q^2$ in the whole
physical region $m_l^2 \leq q^2 \leq {(m_{B_s}- m_{K_0^*})}^2,$ and
using the total mean lifetime $\tau_{B_{s}}= (1.509\pm0.004)~ps$
\cite{PDG}, we present the branching ratio values of semileptonic
decays $B_{s}\to K_0^*\, l\, \bar{\nu}_l$,  ($l=\mu, \tau$) in Table
\ref{T5}, for the three models. Here, we should also stress that the
results obtained for the electron are very close to the results of
the muon, and for this reason, we only present the branching ratios
for the muon in our table. This table contains the results estimated
via the conventional LCSR with the light-meson DA's \cite{YMWang}
and PQCD \cite{RHLi} through S$2$ as well as QCDSR \cite{MZYang}
approaches. Considering the range of errors, the values obtained in
this work are in a logical agreement with the LCSR and PQCD results.
Especially, the obtained values of model II are in a good agreement
with the conventional LCSR. As can be seen in this table,
uncertainties in the values obtained for the branching ratios of the
semileptonic decays $B_{s}\to K_0^*\, l\, \bar{\nu}_l$ are very
large. The main source of errors comes from the form factor
$f_{+}(q^2)$.
\begin{table}[th]
\caption{The branching ratio values of $B_{s}\to K_0^*\, l\,
\bar{\nu}_l$ for the three models and different approaches.}
\label{T5}
\begin{ruledtabular}
\begin{tabular}{cccccccc}
\multirow{3}{*}{\vspace{1.1em}\rm{Mode}} & \multicolumn{3}{c}{This work} &\multirow{3}{*}{\vspace{1.1em} \rm{LCSR (S2)} \cite{YMWang}} &\multirow{3}{*}{\vspace{1.1em} \rm{PQCD (S2)} \cite{RHLi}} & \multirow{3}{*}{\vspace{1.1em}\rm{QCDSR} \cite{MZYang}}   \\
\cline{2-4}                                      & I & II & III &
\\
\hline\\[-2mm]
$\mbox{Br}(B_{s} \to K_0^* \mu \,\nu_{\mu})\times 10^{4}$  &$0.99^{+0.89}_{-0.37}$&$1.67^{+1.32}_{-0.53}$& $1.90^{+1.48}_{-0.63}$  & $1.30^{+1.30}_{-0.40}$ & $2.45^{+1.77}_{-1.05}$ & $0.36^{+0.38}_{-0.24}$ \\[2mm]
$\mbox{Br}(B_{s} \to K_0^* \tau\,\nu_{\tau})\times 10^{4}$ &$0.49^{+0.33}_{-0.17}$&$0.71^{+0.57}_{-0.26}$& $0.65^{+0.55}_{-0.24}$ & $0.52^{+0.57}_{-0.18}$ & $1.09^{+0.82}_{-0.47}$ & $--$                   \\[2mm]
\end{tabular}
\end{ruledtabular}
\end{table}
We show the dependency of the differential branching ratios of
$B_{s}\to K_0^*\, l\, \bar{\nu}_l$, ($l=\mu, \tau$) decays on $q^2$
for the three models in Fig. \ref{F6}.

The semileptonic decays $B_s \to K_0^* l^+ l^-/\nu \bar{\nu}$ are
induced by the FCNC (Appendix). Using the parametrization of these
transitions in terms of the form factors, the differential decay
width in the rest frame of $B_s$-meson can be written as:
\begin{eqnarray}\label{eq31}
\frac{d{\Gamma}}{dq^2}(B_s\rightarrow K_0^* {\nu} \bar \nu) &=& \frac{%
G_{F}^2{\mid}V_{td}V^*_{tb}{\mid}^2
m_{B_s}^3\alpha^2} {2^8 \pi ^5}~\frac{{\mid}D_{\nu}(x_t){\mid}^2}{\rm{sin}^4\theta_W}~\phi%
^{3/2}(1,\hat{r},\hat{s}){\mid}f_+ (q^2){\mid}%
^2~,\nonumber\\
\frac{d\Gamma  }{dq^2}\left( B_s \rightarrow K_0^* l^+l^-\right) &=&
\frac{G_{F}^2{\mid}V_{td}V^*_{tb}{\mid}^2 m_{B_s}^3\alpha^2}
{3\times 2^9\pi ^5}v \,\phi^{1/2}(1,\hat{r},\hat{s})\left[
\left(1+\frac{2\hat{l}} {\hat{s}}\right)
\phi(1,\hat{r},\hat{s})\alpha _1+12~\hat{l}\beta_1\right],
\end{eqnarray}
where $\hat{r}$, $\hat{s}$, $\hat{l}$, $x_t$ and ${\hat m}_b$ and
the functions $v$, $\phi(1,\hat{r},\hat{s})$, $D_{\nu}(x_t)$,
$\alpha_1$ and $\beta_1$ are defined as:
\begin{eqnarray}\label{eq32}
&&\hat{r} = \frac{m_{K_0^*}^2}{m_{B_s}^2}~,\quad\quad \hat{s}
=\frac{q^2}{m_{B_s}^2}~,\quad\quad \hat{l} = \frac{m_{l}^2}{
m_{B_s}^2}~,\quad\quad x_t = \frac{m_t^2}{m_W^2}~,\quad\quad {\hat
m}_b = \frac{m_b}{m_{B_s}}~,\quad\quad
v =\sqrt{1-\frac{4\hat{l}}{\hat{s}}}~,\nonumber\\
&&D_{\nu}(x_t)=\frac{x_t}{8}\Bigg(\frac{2+x_t}{x_t-1}+\frac{3x_t-6}{(x_t-1)^2}\ln
x_t\Bigg)~,\quad\quad\quad\quad {\phi}(1,\hat{r},\hat{s}) =
1+\hat{r}^2
+\hat{s}^2-2\hat{r}-2\hat{s}-2\hat{r}\hat{s}~,\nonumber\\
&&\alpha_1 = \biggl|C_{9}^{\rm eff}\,f_{+}(q^2) +\frac{2\,{\hat
m}_b\,C_{7}^{\rm eff}\,f_{T}(q^2)}{1+\sqrt{\hat{r}}}\biggr|^{2}
+|C_{10}f_{+}(q^2)|^{2} ~,\nonumber\\
&&\beta_1 = |C_{10}|^{2}\biggl[ \biggl( 1+\hat{r}-{\hat{s}\over
2}\biggr) |f_{+}(q^2)|^{2}+\biggl( 1-\hat{r}\biggr) {\rm
Re}(f_{+}(q^2)f_{-}^{*}(q^2))+\frac{1}{2}\hat{s}|f_{-}(q^2)|^{2}\biggr].~
\end{eqnarray}
These expressions contain the Wilson coefficients
$C^{\rm{eff}}_7=-0.313$, $C_9^{\rm{eff}}$ (see Appendix) and
$C_{10}=-4.669$, the CKM matrix elements $|V_{td}V^*_{tb}|=0.008$,
the form factors related to the fit functions, series of functions
and constants. Integrating Eq. (\ref{eq31}) over $q^2$ in the
physical region $4 m_l^2 \leq q^2 \leq {(m_{B_s}- m_{K_0^*})}^2,$
and using $\tau_{B_{s}}$, the branching ratio results of the $B_s
\rightarrow K_0^* l^{+}l^{-}/\nu\bar{\nu}$ are obtained for the
three models as presented in Table \ref{T6}. In this table, we show
only the values obtained by considering the short distance (SD)
effects contributing to the Wilson coefficient $C_9^{\rm eff}$ for
charged lepton case. Predictions by the QCDSR \cite{Ghahramany}, are
smaller than those obtained in this work, because of their estimated
form factors are smaller than ours (see Table \ref{T3}).
\begin{table}[th]
\caption{The branching ratios of the semileptonic $B_s\to K_0^*
l^+l^-/\nu \bar{\nu}$  decays for the three models, including only
the SD effects.} \label{T6}
\begin{ruledtabular}
\begin{tabular}{ccccc}
\multirow{3}{*}{\vspace{1.1em}\rm{Mode}} & \multicolumn{3}{c}{This work} &\multirow{3}{*}{\vspace{1.1em} \rm{QCDSR}\cite{Ghahramany}}  \\
\cline{2-4}                                      & I & II & III &
\\
\hline\\[-2mm]
$\mbox{Br}(B_{s} \to K_0^* \nu\bar{\nu})~~~~\times 10^{7}$ & $0.98^{+0.55}_{-0.27} $&$1.66^{+0.95}_{-0.46}  $&$1.89^{+1.07}_{-0.52}$& $0.25\pm 0.12 $\\[2mm]
$\mbox{Br}(B_{s} \to K_0^* \mu^+ \mu^-)\times 10^{8}$      & $1.32^{+0.75}_{-0.36} $&$2.21^{+1.24}_{-0.62}  $&$2.48^{+1.38}_{-0.69} $& $0.71\pm 0.29 $\\[2mm]
$\mbox{Br}(B_{s} \to K_0^* \tau^+ \tau^-)\times 10^{9}$    & $0.61^{+0.34}_{-0.17} $&$0.63^{+0.35}_{-0.17}  $&$0.45^{+0.25}_{-0.12} $& $0.35\pm 0.16 $\\[2mm]
\end{tabular}
\end{ruledtabular}
\end{table}

\begin{figure}[th]
\begin{center}
\includegraphics[width=7cm,height=2cm]{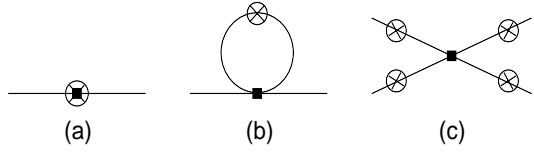}
\caption{Factorizable and nonfactorizable contributions in the LO.
The circled cross marks the possible insertions of the virtual
photon line.} \label{FA1}
\end{center}
\end{figure}

It should be noted that we have computed the branching ratio values
of $B_s\to K_0^* l^+ l^-$ decays in the naive factorization
approximation using the factorizable LO quark-loop, i.e., diagrams
(a) and (b) in Fig. \ref{FA1}. In this method, contributions of the
$O_{1-6}$ operators have the same form factor dependence as $C_9$
which can be absorbed into an effective Wilson coefficient $C^{\rm
eff}_9$.

For a complete analysis of the branching ratio values of $B_s\to
K_0^* l^+ l^-$ decays at the LO, the contributions of the weak
annihilation amplitude of diagram (c) must be added to the form
factor amplitude related to diagrams (a) and (b) in Fig. \ref{FA1}.
Diagram (c) is related to the nonfactorizable effects at the LO.
They arise from electromagnetic corrections to the matrix elements
of purely hadronic operators in the weak effective Hamiltonian.
Since the matrix elements of the semileptonic operators $O_{9,10}$
can be expressed through $B_s\to K_0^*$ form factors,
nonfactorizable corrections contribute to the decay amplitude only
through the production of a virtual photon, which then decays into
the lepton pair \cite{LyZwi,BenFelSei}. These contributions for $B_s
\to K_0^* (1430)$ decays are actually suppressed by small Wilson
coefficients of the penguin operators and can therefore be neglected
in the current analysis. In addition to the factorizable and
nonfactorizable LO diagrams in Fig. \ref{FA1}, there are
factorizable NLO quark-loop and nonfactorizable NLO hard-scattering
and soft-gluon contributions in the FCNC $b \to s$ and $b \to d$
transitions and the effects of them must be taken into account
\cite{KhodjRusov}. Considering the large current uncertainties due
to the form factors, the NLO effects can also be ignored in our
calculations.

\begin{figure}[th]
\begin{center}
\includegraphics[width=5cm,height=5cm]{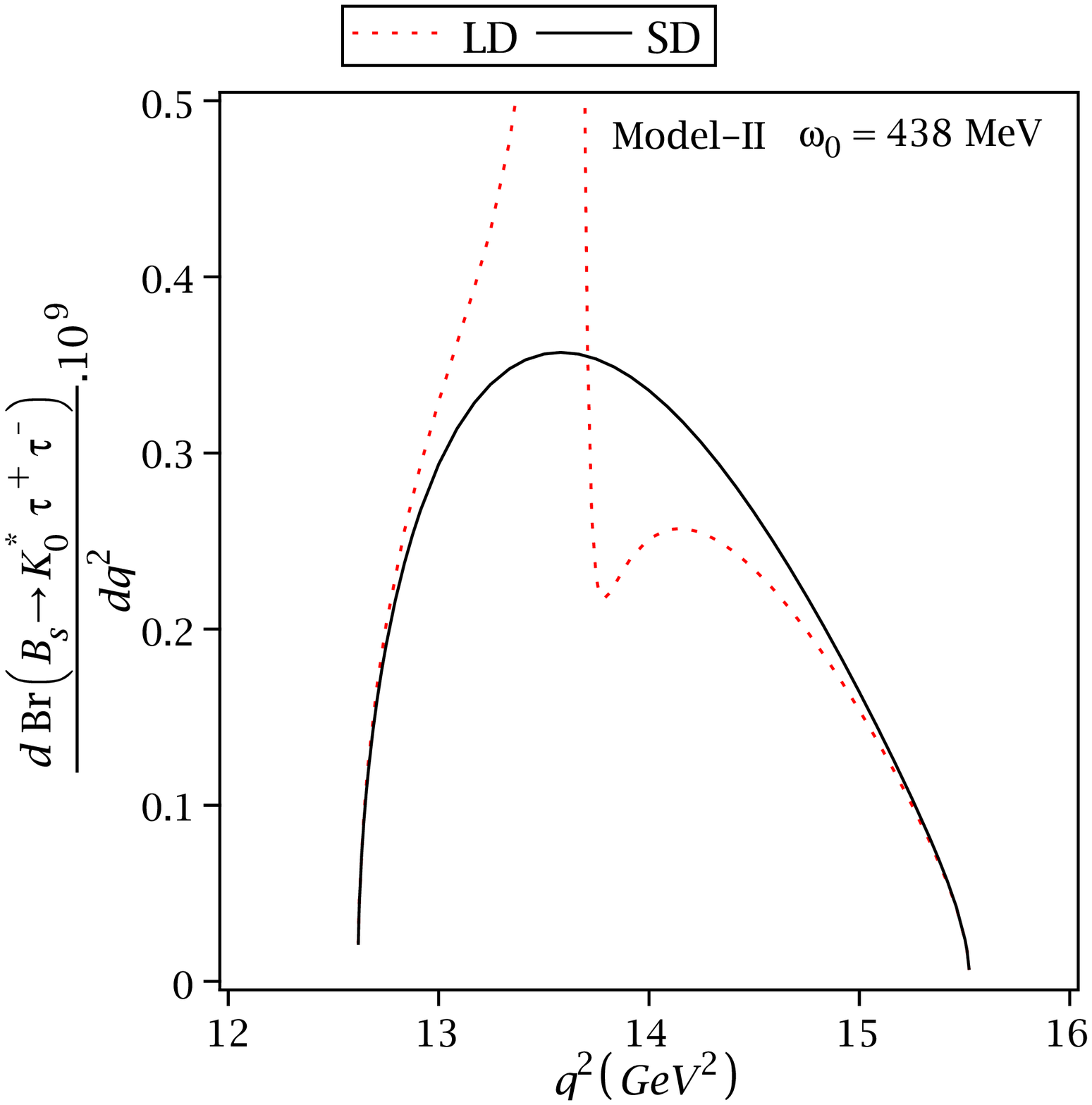}
\includegraphics[width=5cm,height=5cm]{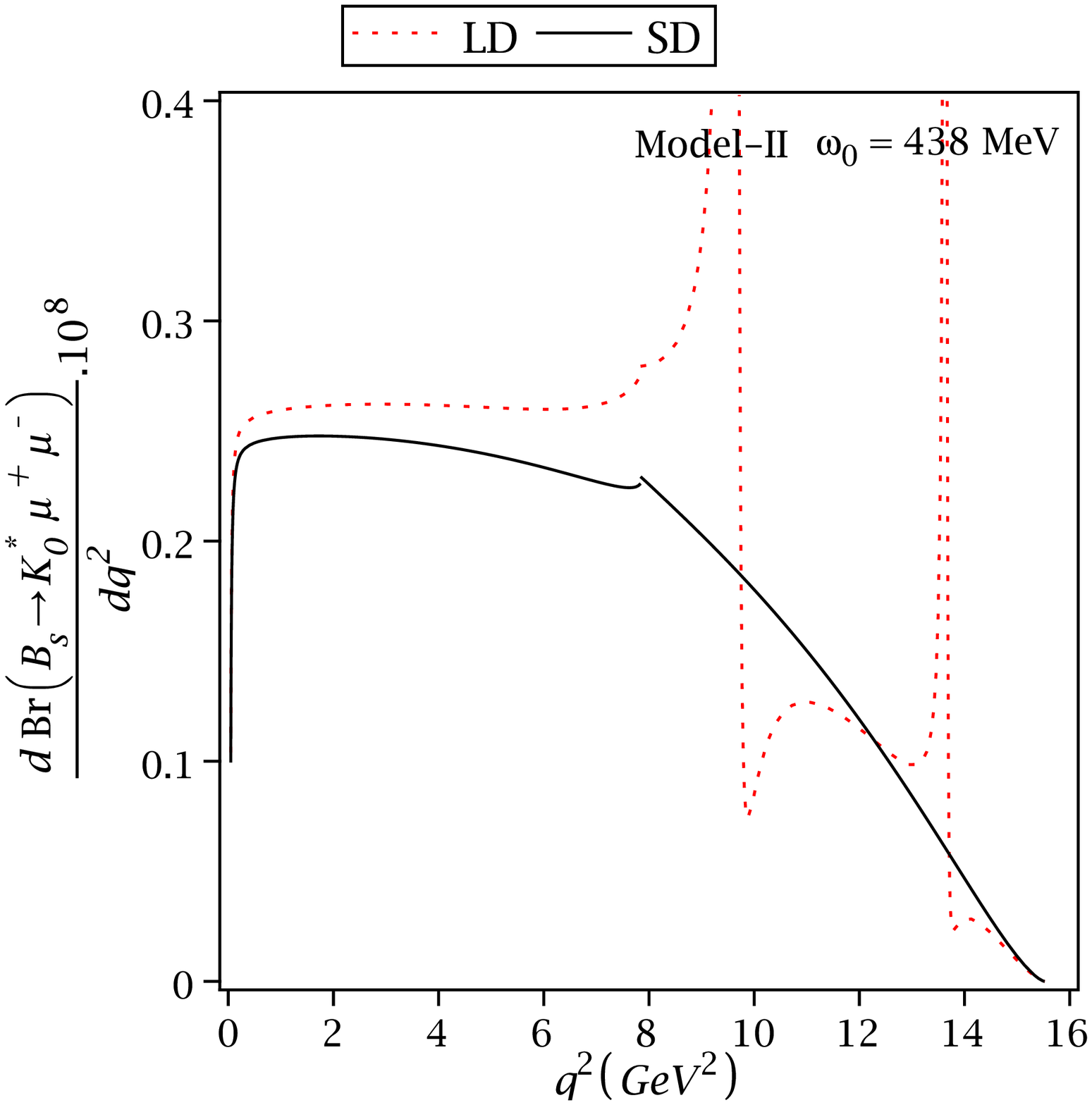}
\includegraphics[width=5cm,height=5cm]{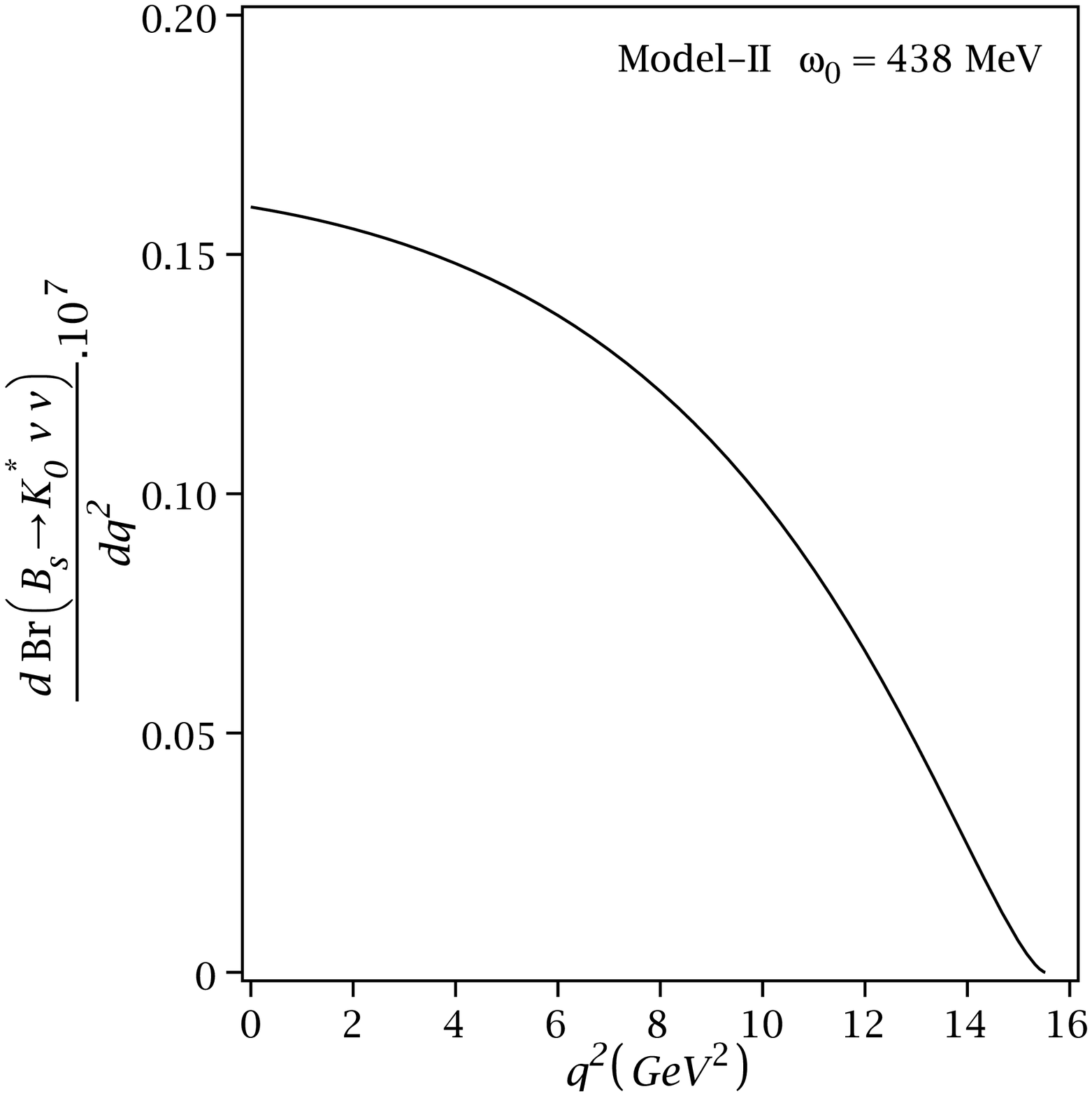}
\caption{The differential branching ratios of the semileptonic $B_s
\to K_0^* l^{+}l^{-}/\nu\bar{\nu}$ decays ($ l=\mu,\tau$)  on $q^2$
for model II.  The solid and dotted lines show the results without
and with the LD effects, respectively.} \label{F7}
\end{center}
\end{figure}

In this part, the branching ratios including LD effects are
presented. In the range of $4m_l^2\leq
q^2\leq(m_{B_s}-m_{K_0^*})^2$, there are two charm-resonances
$J/\psi$ and $\psi(2S)$ used in our calculations. We introduce some
cuts around the resonances of $J/\psi$ and $\psi(2S)$ and study the
following three regions for muon:
\begin{eqnarray}\label{eq33}
\mbox{Region-1}: &&\ \ \ \ \ \ \ \ \sqrt{q^2_{min}} \;\leq\;
\sqrt{q^2} \;\leq\; M_{J/\psi }-0.20,
\nonumber\\
\mbox{Region-2}: && M_{J/\psi}+0.04 \;\leq\; \sqrt{q^2} \;\leq\;
M_{\psi(2S)}-0.10,
\nonumber \\
\mbox{Region-3}: &&\ \ M_{\psi(2S)}+0.02 \;\leq\; \sqrt{q^2}
\;\leq\; m_{B_s}-m_{K_0^*},
\end{eqnarray}
and for tau:
\begin{eqnarray}\label{eq34}
\mbox{Region-2}: & \ \ \ \ \ \ \ \sqrt{q^2_{min}} \;\leq\;
\sqrt{q^2} \;\leq\; M_{\psi(2S)} - 0.02,
\nonumber\\
\mbox{Region-3}: & M_{\psi(2S)} + 0.02\; \leq\; \sqrt{ q^2}\; \leq\;
m_{B_s}-m_{K_0^*},
\end{eqnarray}
where $\sqrt{q^2_{min}}=2m_l$. The branching ratio values for muon
and tau for the three models with LD effects are listed in Table
\ref{T7}.
\begin{table}[th]
\caption{The branching ratios of the semileptonic $B_s\to K_0^*
l^+l^-$  decays for the three models including LD
effects.}\label{T7}
\begin{ruledtabular}
\begin{tabular}{ccccccccccccccccc}
\multirow{3}{*}{\vspace{1.1em}\rm{Mode}} & \multicolumn{3}{c}{$\mbox{Region-1}$} &&\multicolumn{3}{c}{$\mbox{Region-2}$}&&\multicolumn{3}{c}{$\mbox{Region-3}$}&&\multicolumn{3}{c}{$\mbox{Total}$}  \\
\cline{2-4} \cline{6-8}\cline{10-12}\cline{14-16} & I & II & III &&
I & II & III && I & II & III&& I & II & III
\\
\hline\\[-2mm]
$\mbox{Br}(B_s\to K_0^*    \mu^+\mu^-)\times 10^{8}$ &$0.94$&$1.73$&$2.14$&&$0.20$&$0.27$&$0.21$&&$0.02$&$0.02$&$0.01$&&$1.16$&$2.02$&$2.36$\\[2mm]
$\mbox{Br}(B_s\to K_0^*   \tau^+\tau^-)\times 10^{9}$ &$--$&$--$&$--$&&$0.21$&$0.24$&$0.17$&&$0.31$&$0.30$&$0.22$&&$0.52$&$0.54$&$0.39$\\[2mm]
\end{tabular}
\end{ruledtabular}
\end{table}
After numerical analysis, the dependency of the differential
branching ratios for $B_s \to K_0^* l^{+}l^{-}/\nu \bar{\nu}$ on
$q^2$ for model II, with and without LD effects is shown in Fig.
\ref{F7}.

\begin{figure}[th]
\begin{center}
\includegraphics[width=6cm,height=6cm]{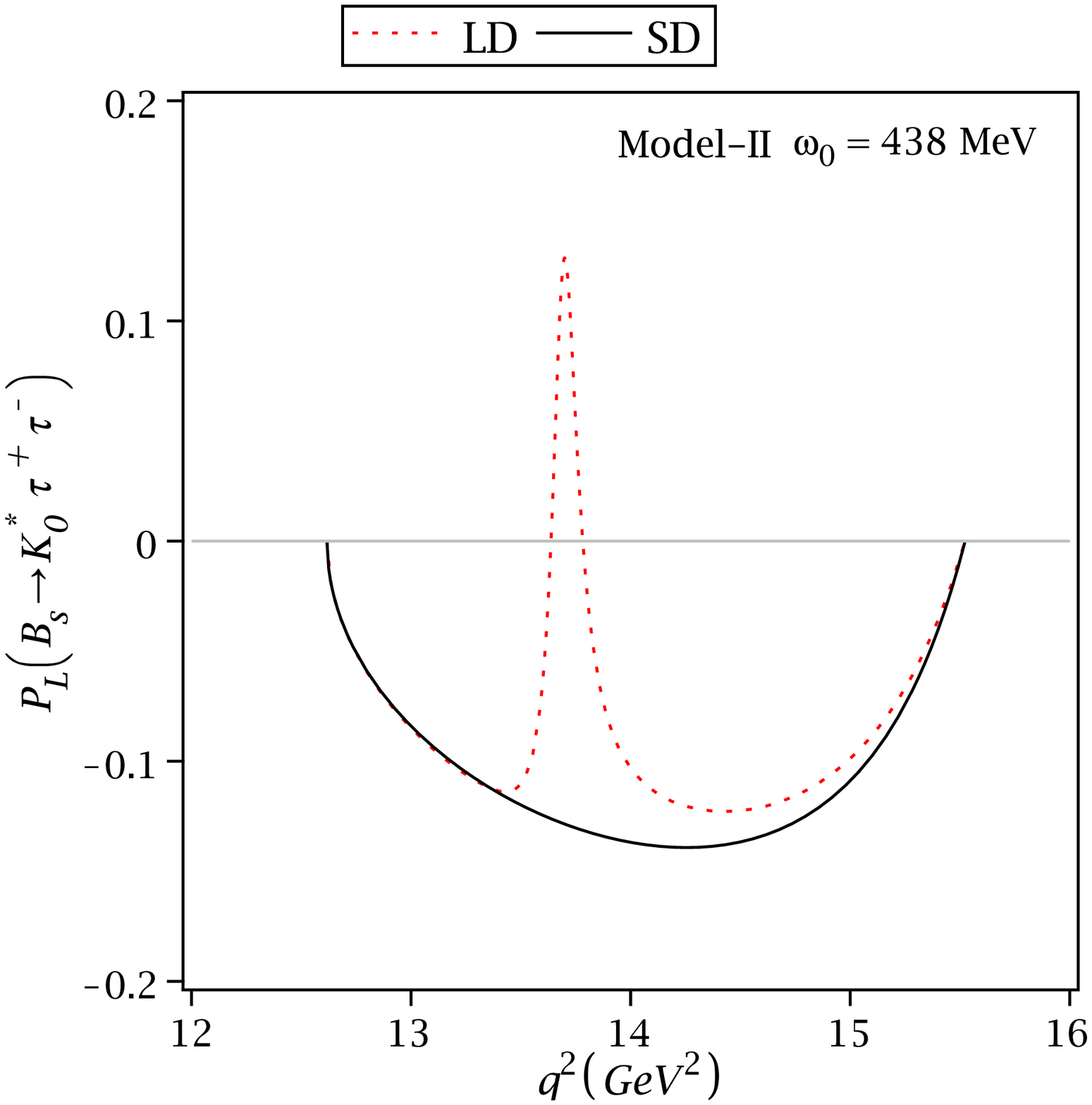}
\includegraphics[width=6cm,height=6cm]{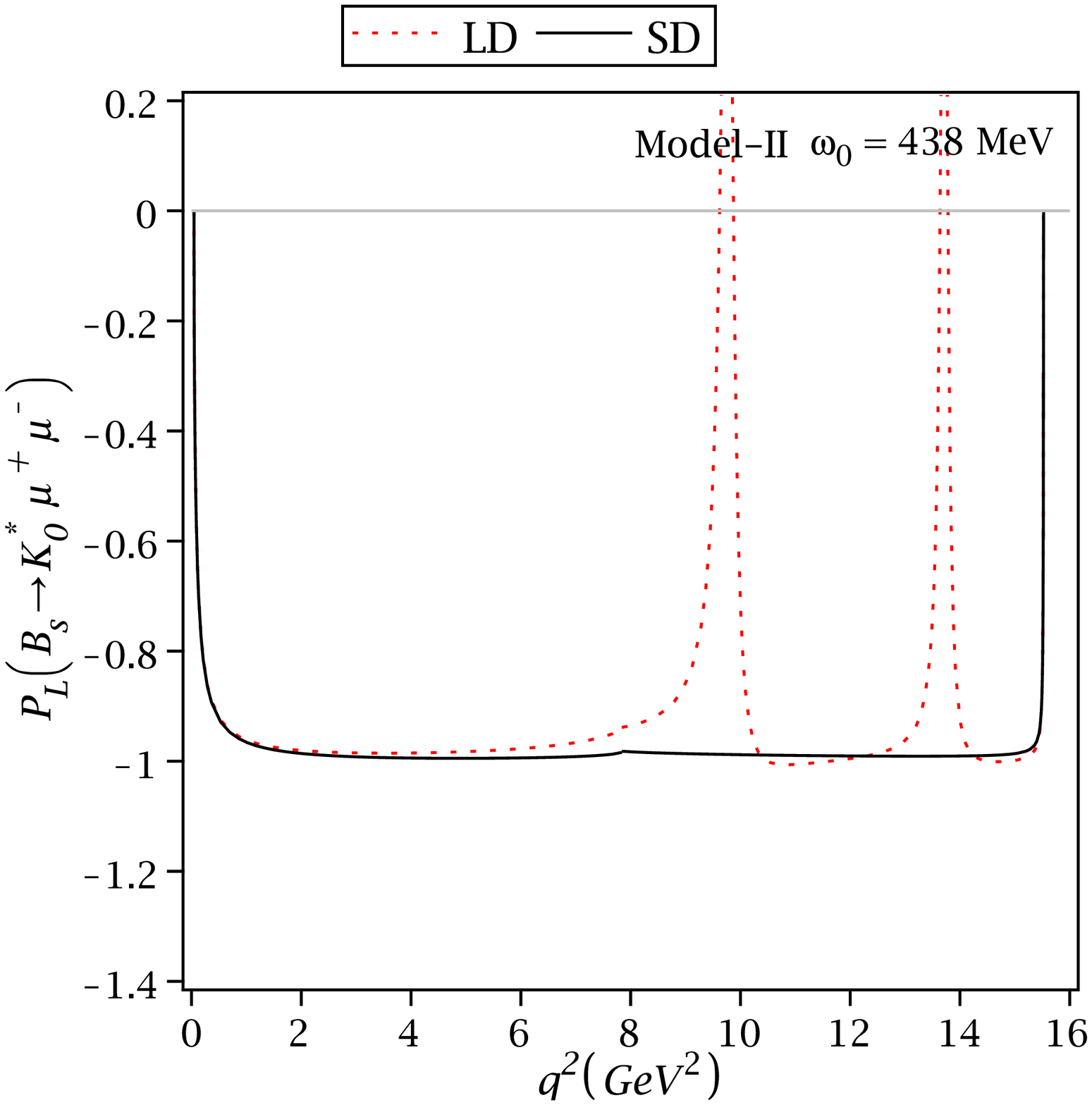}
\caption{The dependence of the longitudinal lepton polarization
asymmetries on $q^2$ for model II. The solid and dotted lines show
the results without and with the LD effects, respectively.}
\label{F8}
\end{center}
\end{figure}

Finally, we want to calculate the longitudinal lepton polarization
asymmetries for the considered decays. The longitudinal lepton
polarization asymmetry formula for $B_s \to K_0^* l^{+}l^{-}$ is
given as:
\begin{eqnarray}\label{eq35}
P_L=\frac{2v}{(1+\frac{2\hat{l}}{\hat{s}})\phi(1,\hat{r},\hat{s})\alpha_1+12
\hat{l}\beta_1}{\rm{Re}}\left[\phi(1,\hat{r},\hat{s})\left(C_9^{eff}f_+(q^2)-\frac{2C_7
f_T(q^2)}{1+\sqrt{\hat{r}}} \right)(C_{10}f_+(q^2))^*\right],
\end{eqnarray}
where $v, ~\hat{l}, ~\hat{r}, ~\hat{s}, ~\phi(1,\hat{r},\hat{s}),
\alpha_1$ and $\beta_1$ were defined before. The dependence of the
longitudinal lepton polarization asymmetries for the $B_s \to K_0^*
l^{+}l^{-},~(l=\mu, \tau)$ decays on the transferred momentum square
$q^2$ for model II, with and without LD effects is plotted in Fig.
\ref{F8}.
The averaged values of the lepton polarization asymmetries of these
decays for the three models, without the LD contributions are
obtained and presented in Table \ref{T8}. These polarization
asymmetries provide valuable information on the flavor changing loop
effects in the SM.
\begin{table}[th]
\caption{Averaged values of the lepton polarization asymmetries of
$B_s \to K_0^* l^{+}l^{-},~(l=\mu, \tau)$ decays for the three
models, without the LD contributions.}\label{T8}
\begin{ruledtabular}
\begin{tabular}{cccc}
\rm{Model} & \rm I & \rm II & \rm III\\
\hline\\[-2mm]
$\langle P_{L}\rangle_{\mu}$ &$ -0.80$&$ -0.88$&$ -0.98$\\[2mm]
$\langle P_{L}\rangle_{\tau}$&$ -0.12$&$-0.13$&$-0.11$\\[2mm]
\end{tabular}
\end{ruledtabular}
\end{table}

\section{Conclusion}\label{sec4}

In summary, the transition  form factors of the semileptonic
$B_{s}\to K_{0}$  transitions were calculated via the LCSR with the
$B_s$-meson DA's in the $\rm SU(3)_F$ symmetry limit. We considered
the three different models for the shapes of the two-particle DA's,
$\varphi_{\pm}$. It was shown that in estimation of the form
factors, the main uncertainties came from the shape parameter
$\omega_{0}$ and the decay constant of the $K_0^*$-meson. In this
work, we used  $\omega_{0}=\lambda_{B_s}$. Recently, the inverse
moment of the $B_s$-meson distribution amplitude, $\lambda_{B_s}$
has been predicted from the QCDSR method as $\lambda_{B_s}=(438 \pm
150)~\rm{MeV}$. There was a very good agreement between our results
for the form factors at zero momentum transfer in model II and
predictions of the conventional LCSR with the light-meson DA's in
scenario 2. Therefore, our calculations confirmed scenario 2 for
describing the scalar meson $K_0^*(1430)$. Using the form factors
$f_{+}(q^2)$, $f_{-}(q^2)$ and $f_{T}(q^2)$, the branching ratio
values for the semileptonic $B_s \to K_0^* l \bar{\nu_{l}}$ and $B_s
\to K_0^* l \bar{l}/\nu \bar{\nu}$ ($l=e, \mu, \tau$) decays were
calculated. It is worth mentioning that we computed the branching
ratio values of $B_s\to K_0^* l^+ l^-$ decays in the naive
factorization approximation. Considering the SD and LD effects, the
dependence of the differential branching ratios as well as the
longitudinal lepton polarization asymmetries for $B_s \to K_0^* l
\bar{l}$ decays were investigated with respect to $q^2$. Future
experimental measurement can give valuable information about these
aforesaid decays and the nature of the scalar meson $K_0^*(1430)$.

\clearpage
\appendix
\begin{center}
{\Large \textbf{Appendix: The effective weak Hamiltonian of the $b
\to d~ l^+ l^-$ transition}}
\end{center}

The effective weak Hamiltonian of the $b \to d~ l^+ l^-$ transition
has the following form in the SM:
\begin{eqnarray*}\label{A1}
\mathcal{H}^{b\to d}_{\rm eff} = - \frac{ G_F}{\sqrt{2}}\left(
V_{ub}V_{ud}^{*}\sum_{i=1}^{2} C_i(\mu)
O^{u}_i(\mu)+V_{cb}V_{cd}^{*}\sum_{i=1}^{2} C_i(\mu)
O^{c}_i(\mu)-V_{tb}V_{td}^{*}\sum_{i=3}^{10} C_i(\mu)
O_i(\mu)\right),
\end{eqnarray*}
where $V_{jk}$ and $C_i(\mu)$ are the  CKM matrix elements and
Wilson coefficients, respectively. The local operators are
current-current operators $O^{u,c}_{1,2}$, QCD penguin operators
$O_{3-6}$, magnetic penguin operators $O_{7,8}$, and semileptonic
electroweak penguin operators $O_{9,10}$. The explicit expressions
of these operators  for $b \to d l^+ l^-$ transition are written as
\cite{Buras0}
\begin{eqnarray*}\label{A2}
\begin{array}{ll}
O_1     =  (\bar{d}_{i}  c_{j})_{V-A} , (\bar{c}_{j}  b_{i})_{V-A}
,               &
O_2     =  (\bar{d} c)_{V-A}  (\bar{c} b)_{V-A}   ,              \\
O_3     =  (\bar{d} b)_{V-A}\sum_q(\bar{q}q)_{V-A}  ,            &
O_4     =  (\bar{d}_{i}  b_{j})_{V-A} \sum_q (\bar{q}_{j}
q_{i})_{V-A} ,                                    \\
O_5     =  (\bar{d} b)_{V-A}\sum_q(\bar{q}q)_{V+A} ,            &
O_6     =  (\bar{d}_{i}  b_{j })_{V-A}
\sum_q  (\bar{q}_{j}  q_{i})_{V+A}  ,               \\
O_7     =  \frac{e}{8\pi^2} m_b (\bar{d} \sigma^{\mu\nu}
(1+\gamma_5) b) F_{\mu\nu}  ,                     & O_8    =
\frac{g}{8\pi^2} m_b (\bar{d}_i \sigma^{\mu\nu}
(1+\gamma_5) { T}_{ij} b_j) { G}_{\mu\nu}  ,          \\
O_9     = \frac{e}{8\pi^2} (\bar{d} b)_{V-A}  (\bar{l}l)_V ,      &
O_{10}  = \frac{e}{8\pi^2} (\bar{d} b)_{V-A}  (\bar{l}l)_A ,
\end{array}
\end{eqnarray*}
where ${ G}_{\mu\nu}$ and $F_{\mu\nu}$ are the gluon and photon
field strengths, respectively; ${T}_{ij}$ are the generators of the
$SU(3)$ color group; $i$ and $j$ denote color indices. Labels $(V\pm
A)$ stand for $\gamma^\mu(1\pm\gamma^5)$. The magnetic and
electroweak penguin operators $O_{7}$, and $O_{9,10}$ are
responsible for the SD effects in the FCNC $b \to d$ transition, but
the operators $O_{1-6}$  involve both SD and LD contributions in
this transition. In the naive factorization approximation,
contributions of the $O_{1-6}$ operators have the same form factor
dependence as $C_9$ which can be absorbed into an effective Wilson
coefficient $C^{\rm eff}_9$. The effective Wilson coefficient
$C_{9}^{\rm{eff}}$ includes both the SD and LD effects as
\begin{eqnarray*}\label{A3}
C^{\rm eff}_9 = C_9 + Y_{SD}(q^2)+Y_{LD}(q^2),
\end{eqnarray*}
where $Y_{SD}(q^2)$ describes the SD contributions from four-quark
operators far away from the resonance regions, which can be
calculated reliably in perturbative theory as \cite{Buras0,Aliev0}:
\begin{eqnarray*}\label{A4}
Y_{SD}(q^2)&=&0.138~
\omega(s)+h(\hat{m_c},s)C_0+\lambda_u\,h(\hat{m_c},s)(3C_1+C_2)
-\frac{1}{2}h(1,s)(4C_3+4C_4+3C_5+C_6)\nonumber\\
&-&\frac{1}{2}h(0,s)(2\lambda_u [3C_1+C_2]+C_3+3C_4)
+\frac{2}{9}(3C_3+C_4+3C_5+C_6),
\end{eqnarray*}
where $s=q^2/m_b^2$, $\hat{m_c}=m_c/m_b$,
$C_0=3C_1+C_2+3C_3+C_4+3C_5+C_6$, $\lambda_u
=\frac{V_{ub}V^*_{ud}}{V_{tb}V^*_{td}}$,    and
\begin{eqnarray*}\label{A5}
\omega(s)= -\frac{2}{9} \pi^2 -\frac{4}{3} {\rm Li}_2(s)-
\frac{2}{3} \ln (s) \ln(1-s) - \frac{5+4s}{3(1+2s)} \ln(1-s) -
\frac{2 s(1+s)(1-2s)}{3(1-s)^2(1+2s)} \ln (s)+ \frac{5+9 s-6
s^2}{3(1-s)(1+2s)},
\end{eqnarray*}
represents the ${\cal O}(\alpha_s)$ correction coming from one gluon
exchange in the matrix element of the operator $O_9$ \cite{Jezabek},
while  $h(\hat m_c,  s)$ and $h(0,  s)$ represent one-loop
corrections to the four-quark operators $O_{1-6}$ \cite{Misiak}. The
functional form of the $h(\hat{m_c}, s)$ and $h(0, s)$ are as:
\begin{eqnarray*}\label{A6}
h(\hat m_c,  s)=- \frac{8}{9}\ln \frac{m_b}{\mu} -
\frac{8}{9}\ln\hat m_c + \frac{8}{27} + \frac{4}{9} x -\frac{2}{9}
(2+x) |1-x|^{1/2} \left\{
\begin{array}{ll}
\left( \ln\left| \frac{\sqrt{1-x} + 1}{\sqrt{1-x} - 1}\right| - i\pi
\right), &
\mbox{for } x \equiv \frac{4 \hat m_c^2}{ s} < 1 \nonumber \\
& \\
2 \arctan \frac{1}{\sqrt{x-1}}, & \mbox{for } x \equiv \frac {4 \hat
m_c^2}{ s} > 1
\end{array}
\right. \nonumber\\
\end{eqnarray*}
and
\begin{eqnarray*}\label{A7}
h(0,  s)  =  \frac{8}{27} -\frac{8}{9} \ln\frac{m_b}{\mu} -
\frac{4}{9} \ln\ s + \frac{4}{9} i\pi\,.
\end{eqnarray*}
The LD contributions, $Y_{LD}(q^2)$ from four-quark operators near
the $u\bar{u}$, $d\bar{d}$ and $c\bar{c}$ resonances cannot be
calculated from the first principles of QCD and are usually
parametrized in the form of a phenomenological Breit-Wigner formula
as \cite{Buras0,Aliev0}:
\begin{eqnarray*}
Y_{LD}(q^2)&=&\frac{3\pi}{\alpha^2}
\left\{\left(C_0+\lambda_u[3C_1+C_2]\right)\sum_{V_i=J/\psi,
\psi(2S)}\frac{\Gamma(V_i\to l^+ l^-)m_{V_i}}{m_{V_i}^2-q^2-i
m_{V_i} \Gamma_{V_i}} \right\}.
\end{eqnarray*}
In the range of $4m_l^2\leq q^2\leq(m_{B_s}-m_{K_0^*})^2$, there are
two charm-resonances $J/\psi(3.097)$ and $\psi(3.686)$ used in our
calculations.

\end{document}